\documentclass[a4paper,10pt]{svmult}
\usepackage[utf8x]{inputenc}
\usepackage{graphicx}

\usepackage{multicol}        
\usepackage[bottom]{footmisc}

\begin{document}
\title*{
 First-Principles Study of the 
 Electronic and Magnetic Properties of Defects in Carbon Nanostructures}

\titlerunning{First-Principles Study of Defects in Carbon Nanostructures}

\author{
Elton J. G. Santos, Andr\'es Ayuela and Daniel S\'anchez-Portal}

\institute{
Elton  J. G. Santos$^{\dagger}$, Andr\'es Ayuela$^{\star}$ and Daniel 
S\'anchez-Portal$^{\S}$ \at
Centro de F\'{\i}sica de Materiales (CFM-MPC) CSIC-UPV/EHU, 
Paseo Manuel de Lardizabal 5, 20018 San Sebasti\'an, Spain, and \\
Donostia International Physics Center (DIPC), Paseo Manuel de Lardizabal 4, 
20018 San Sebasti\'an, Spain. \\
$^\star$\email{swxayfea@ehu.es} \\
$^\S$\email{sqbsapod@ehu.es} \and 
$^{\dagger}$Present address: School of Engineering and 
Applied Sciences, Harvard University, Cambridge, Massachusetts 02138, USA.\\
\email{esantos@seas.harvard.edu}.
}

\maketitle

\abstract{
Understanding the 
magnetic properties of graphenic nanostructures is instrumental 
in future spintronics applications. These magnetic 
properties are known to 
depend crucially 
on the presence of defects.
Here we 
review our
recent theoretical studies using density functional 
calculations on two types
of defects in carbon nanostructures: 
Substitutional doping with transition metals,  and sp$^3$-type defects created
by covalent functionalization with organic and inorganic molecules. 
We focus on such defects because they
can be used to create and control magnetism in graphene-based materials. 
Our main results are summarized as follows:
\begin{enumerate} 
\item[({\it i})]Substitutional metal impurities are fully understood using a 
          model based on the hybridization between the $d$ states of the metal atom
          and the defect levels associated with an unreconstructed 
          D$_{3h}$ carbon vacancy.  We identify three different regimes, associated with  the 
          occupation of 
distinct
         hybridization levels, which determine
        the magnetic properties obtained 
        with this type of doping;
\item[({\it ii})] A spin moment of 1.0~$\mu_B$ is {\it always} induced 
by chemical functionalization 
when a molecule chemisorbs on a graphene 
layer via a single C-C (or other weakly polar) covalent bond. 
The magnetic coupling between adsorbates shows 
a key dependence on the sublattice adsorption site.
This effect is similar to that of H adsorption, 
however,
with universal character;
\item[({\it iii})] The spin moment of substitutional metal 
impurities can be controlled using strain.
In particular, we show that although Ni substitutionals 
are non-magnetic in flat and unstrained
graphene, the magnetism of these
defects can be activated by applying either uniaxial strain or curvature
to the graphene layer.  
\end{enumerate}
All these results provide key information about formation and control of defect-induced magnetism in graphene and related materials.
}

\section{Introduction}
\label{intro}

%

The experimental 
discovery
of graphene, a truly two-dimensional crystal, 
has led to the rapid development of a very active line of research.
Graphene is not only a fundamental model to study other 
types of carbon materials, but exhibits many uncommon electronic
properties governed by a Dirac-like wave 
equation~(Novoselov et al. 2004 and 
2005) 
Graphene, which exhibits ballistic electron
transport on the submicrometre scale, is considered a key
material for the next generation of carbon-based electronic 
devices~(Geim et al 2007,  Castro Neto et al. 2009).  
In particular, 
carbon-based
materials are quite promising
for spintronics and related
applications due to their
long spin relaxation and
decoherence times owing to the 
spin-orbit interaction
and the 
hyperfine interaction of the electron spins with the
carbon nuclei, 
both negligible
~(Hueso et al. 2008, Trauzettel et al. 2007, Tombros et al. 2007, Yazyev 2008).
In addition, the possibility to control
the magnetism of edge states in nanoribbons and nanotubes
by applying external electric fields introduces an additional
degree of freedom to control
the spin transport~(Son et al. 2006, Ma{\~n}anes et al. 2008).
Nevertheless, for the design of realistic 
devices,
the effect
of defects and impurities 
has to be taken into account.
Indeed, 
a substantial amount 
of work has been devoted
to the study of defects and different types of impurities in these
materials. 
The magnetic properties of point defects, like vacancies, adatoms or
substitutionals, have been 
recognized by many authors~(Lehtinen et al. 2003, Palacios et al. 2008, 
Kumazaki et al. 2008, Santos et al. 2010a, Santos et al. 2010b, 
Fuhrer et al. 2010, Ugeda et al. 2010, Yazyev et al. 2007). 
It has now become clear that the presence of defects can affect the
operation of graphene based devices and can be used
to tune their response. 

In the present 
Chapter,
we provide a review of some 
recent computational studies 
on the 
role of some particular type of defects 
in determining 
the electronic and, in particular, the magnetic 
properties of graphene and carbon nanotubes (Santos et al. 2008, 2010a, 2010b, 2010c, 2011, 2012a, 2012b). 
We will consider two types of defects: Substitutional transition metals and
covalently bonded adsorbates. 
For
the substitutional transition metals, we first present 
some of the existing experimental evidence about the 
presence of such impurities in graphene and carbon 
nanotubes.
Then, we summarize our results for the structural, electronic and magnetic properties
of substitutional transition metal impurities in graphene. 
We show that all these properties can be explained using a simple 
model based on hybridization between the $d$ shell of the metal 
atoms and the defects states of an unrelaxed carbon vacancy. 
We also show that it is possible to change the local spin moment
of the substitutional impurities by applying mechanical deformations
to the carbon layer. This effect is studied in detail in the case of
Ni substitutionals. Although these impurities are non-magnetic at 
a
zero 
strain, we demonstrate that it is possible to switch on the
magnetism of Ni-doped graphene either by applying uniaxial 
strain or curvature to the carbon layer. 
Subsequently, 
we explore the magnetic properties induced by covalent functionalization
of  graphene and carbon nanotubes. 
We find that the magnetic properties in this case are universal, in the 
sense that they are largely independent
of the particular adsorbate: As far the adsorbate is attached to the carbon layer
through a single C-C covalent bond (or other weakly-polar covalent bond), there
is always a spin moment of 1~$\mu_B$ associated with each adsorbate.
We show that this result
can be understood in terms of a simple model based on the 
so-called $\pi-$vacancy, 
i.e. one $p_z$ orbital removed from a $\pi$-tight-binding description of graphene. 
This model captures the main features induced by the covalent functionalization 
and the physics 
behind.
In particular, using this model we can easily predict the total spin moment
of the system when there are several molecules attached to the carbon layer
simultaneously.  Finally, we have 
also studied in detail the magnetic couplings
between Co substitutional impurities in graphene. Surprisingly, the Co substitutional
impurity is also well described in terms of a simple $\pi$-vacancy model.

\section{
Substitutional Transition Metal Impurities in Graphene}
\label{substitutionals}

\subsection{Experimental Evidences}
\label{substitutionals-exp}

Direct experimental evidence of the existence of substitutional impurities in graphene, 
in which a single metal atom substitutes one or several carbon atoms in the layer,
has been recently provided by Gan and coworkers (Gan et al. 2008).
Using high-resolution transmission electron microscopy 
(HRTEM), 
these authors 
could
visualize individual
Au and Pt atoms incorporated into  a very thin graphitic layer probably
consisting of one or two graphene 
layers.
From the real-time evolution and temperature dependence of the
dynamics,
they obtained information
about the diffusion of these atoms. 
Substantial
diffusion
barriers ($\sim$2.5~eV) were observed for in-plane migration, which
indicates the large
stability of these defects and the presence
of strong carbon-metal bonds. These observations
indicate that the atoms occupy substitutional
positions. 

In another experiment using 
double-walled carbon nanotubes (DWCNT) (Rodriguez-Manzo et al. 2010), 
Fe atoms were trapped at vacancies likewise 
the previous observations for graphene layers. 
In these experiments, 
the electron beam was directed onto a predefined position and kept stationary for few seconds
in order to create a lattice defect. 
Fe atoms 
had been previously deposited on the nanotube 
surface before the defect formation.
After irradiation, a bright spot in the dark-field image was observed. A quantitative 
analysis of the intensity profile showed an increase of the scattered intensity at the irradiated 
position relative to the center of the pristine DWCNT. This demonstrates that at the defect position, on the 
top or bottom side of the 
DWCNT, a Fe atom was trapped.

Recent evidence 
was also 
reported for substitutional Ni impurities in
single-walled carbon nanotubes (SWCNT)~(Ushiro et al. 2006)
and graphitic particles~(Banhart et al. 2000).
Ushiro and co-authors (Ushiro et al. 2006) showed that Ni
substitutional defects were present in SWCNT samples 
synthesized using Ni-containing catalysts 
even after careful purification. According to their analysis of 
X-ray absorption data (XANES), the most
likely configuration 
for these defects
has a Ni atom replacing a carbon atom.

The presence of substitutional defects 
in the samples 
can have important implications for the interpretation of some experiments.
For example, substitutional
atoms
of 
transition metals are
expected to strongly influence the magnetic properties of graphenic
nanostructures.  Interestingly, transition metals like Fe, Ni or Co
are among the most common catalysts used for the production of
SWCNTs~(Dresselhaus et al. 2001). Furthermore, the experiments by 
Banhart
and coworkers~(Rodriguez-Manzo et al. 2009) have demonstrated 
that it is possible
to create individual vacancies at desired locations in carbon
nanotubes using electron beams. This experiment, 
combined with
the observed stability of substitutional impurities, opens a
route to fabricate new devices incorporating substitutional
impurities at 
predefined 
locations. These
devices would allow for the experimental verification of the
unusual magnetic interactions mediated by the graphenic carbon
network that have been 
predicted recently~(Brey et al. 2007, Kirwan et al. 2008, Santos et al. 2010a).
This becomes particularly interesting in the light of the
recent finding that the spin moment of 
that
impurities could 
be easily tuned by applying uniaxial strain and/or mechanical 
deformations to the carbon layer~(Santos et al. 2008, Huang et al. 2011, Santos et al. 2012a).
In spite of this, the magnetic properties of substitutional transition-metal 
impurities in graphenic systems 
were not
studied in detail 
until very recently.
Few calculations have considered the effect of this kind of doping on
the 
magnetic properties of the graphenic 
materials,
and this will be 
the main topic
of the following sections. 

\subsection{Structure and Binding 
}
\label{substitutionals-theory1}

\begin{figure}\centering
\includegraphics[width=3.250in]{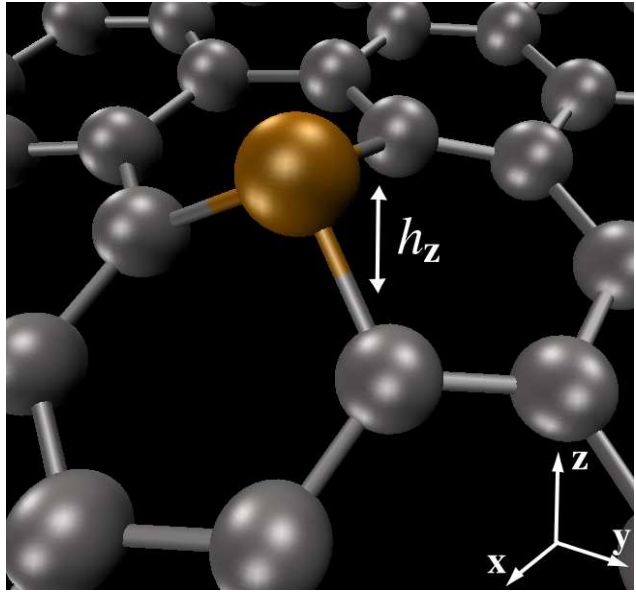}
\caption{\label{fig:fig1-njp} Typical geometry of transition and noble
substitutional metal atoms in graphene. The metal atom moves upwards
from the layer and occupies, in most cases, an almost perfectly symmetric
three-fold position with C$_{3v}$ symmetry.
}
\end{figure}

\begin{figure}\centering
\includegraphics[width=3.250in]{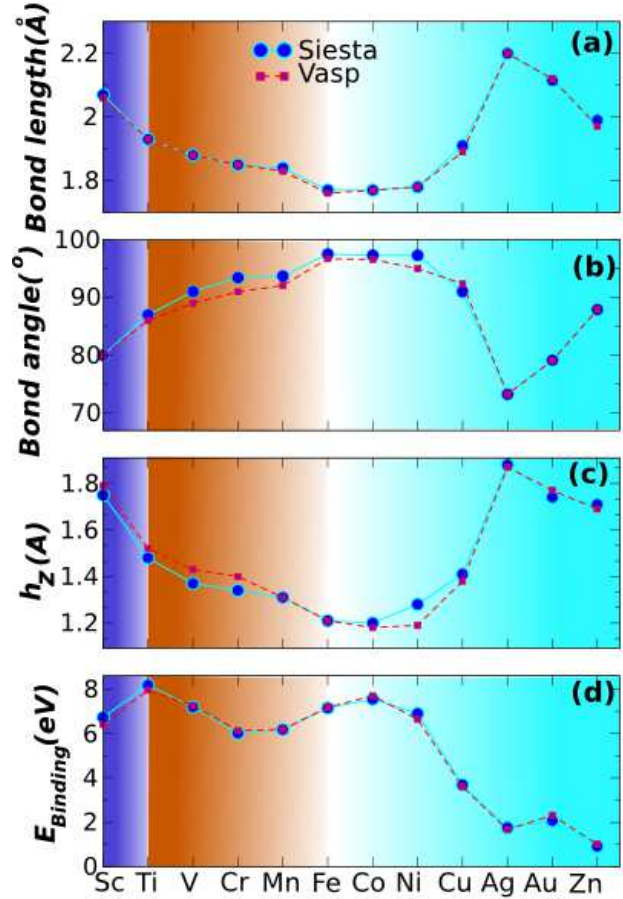}
\caption{\label{fig:fig2-njp}
Structural parameters and binding energies of
substitutional transition and noble metals in graphene.
Bond lengths and angles have been averaged for the noble metals.
The data presented for Zn correspond to the high-spin
solution with C$_{3v}$ symmetry, and are very close to the
averaged results for the most stable distorted solution. Adapted from (Santos et al. 2010b).
}
\end{figure}

In Fig.~\ref{fig:fig1-njp} we show the typical geometry
found in our calculations for 
a graphene layer where one carbon atom has been substituted by a metal impurity.  
The metal atom appears always displaced from the carbon layer. The height over the
plane defined by its three nearest carbon neighbors is
in the range 1.7-0.9~\AA. These three carbon atoms are also
displaced over the average position of the graphene layer by 0.3-0.5~\AA.
The total height (h$_z$) of the metal atom over the graphene plane
is the sum of these two contributions and ranges between 1.2-1.8~\AA, as shown in panel (c) of
Fig.~\ref{fig:fig2-njp}.

In most cases the metal atom occupies an almost perfectly symmetric
configuration with  C$_{3v}$ symmetry. Exceptions are the 
studied
noble metals,
that are slightly displaced from the central position, and Zn that suffers a Jahn-Teller
distortion in its most stable configuration. However, we have found that
it is also possible to stabilize a symmetric configuration for Zn with a
binding energy only $\sim$150~meV smaller. This configuration 
will be referred to 
as Zn$_{{\rm C}_{3v}}$ throughout 
this chapter.

Figures~\ref{fig:fig2-njp}(a)-(c) present a summary of the
structural parameters of substitutional 3$d$ transition metals, noble metals and Zn
in graphene. 
Our calculations are in very good agreement with the
results of a similar study performed by Krasheninnikov et al.~(Krasheninnikov et al. 2009),
although they overlooked the existence of the high-spin Zn$_{{\rm C}_{3v}}$ configuration. 
Solid circles correspond to calculations using the 
{\sc Siesta} code~(Soler et al. 2002) 
with pseudopotentials~(Troullier and Martins 1993) 
and a basis set of atomic orbitals (LCAO), while open squares stand for
{\sc Vasp}~(Kresse and Hafner 1993, Kresse and Furthm{\"u}ller 1996) 
calculations using plane-waves and 
PAW potentials~(Bl{\"o}chl 1994). 
As we can see,
the agreement between both sets of calculations is excellent. Data in these figures correspond
to calculations using a 4$\times$4 supercell of graphene. For several metals we have also performed calculations
using a larger 8$\times$8 supercell and find almost identical
results. This is particularly true for the total spin moments, which are 
less dependent on the size of the supercell, but require a sufficiently dense
k-point sampling to converge. 
The behavior of the metal-carbon bond length and h$_z$
reflect approximately the size of the metal atom. 
For transition metals these distances decrease as we increase the atomic number,
with a small discontinuity when going from Mn to Fe. The
carbon-metal bond length reaches its minimum for Fe
(d$_{\rm{C-Fe}}$=1.76~\AA), keeping a very similar value for Co and
Ni. For Cu and Zn the distances increase reflecting the fully
occupied 3$d$ shell and the large size of the 4$s$ orbitals. Among
the noble metals we find that, as expected, the bond length largely
increases for Ag with respect to Cu, but slightly decreases when
going from Ag to Au. The latter behavior 
is
understood from the
compression of the 6$s$ shell due to scalar relativistic effects.

\begin{table}\centering
\caption{\label{tab:geometals}
Structural parameters for substitutional
noble metals 
and Zn 
in graphene.
d$_{\rm{C-M}}$ indicates the bond distances between the
metal atom and its three carbon neighbors and h$_z$ is the height
of the impurity over the carbon layer (see the text). The bond angles
are also given. 
}

\begin{tabular}{@{}lccc}
\hline \\[-1ex]
& d$_{\rm{C-M}}$(\AA) & h$_z$(\AA)
& $\theta$ ($^{\circ}$)  \\
\hline \\[-1ex]
Cu &  1.93, 1.90, 1.90
  & 1.40
   & 88.9, 88.9, 95.2  \\

 Ag &  2.23, 2.19, 2.19
     & 1.84
      &  71.7, 71.7, 76.7    \\
 Au & 2.09, 2.12, 2.12
  &   1.71
      & 78.0, 78.0, 81.6  \\
Zn &  2.06, 1.89, 1,89
     & 1.54
      &  88.3 88.3 103.9    \\
Zn$_{{\rm C}_{3v}}$  & 1.99
  & 1.67
   & 87.9  \\
\hline \\[-1ex]
\end{tabular}
\end{table}

As already mentioned, noble metals and Zn present a distorted configuration.
In Table~\ref{tab:geometals}
we find the corresponding structural parameters.
For Cu and Ag one of the metal-carbon bond lengths
is slightly larger than the other two, whereas for Au one is shorter than the others. However, the
distortions are rather small with variations of the bond lengths smaller than 2~\%.
The larger scalar relativistic effects in Au give rise to slightly 
smaller metal-carbon bond lengths for this metal as compared to Ag.
In the case of Zn atoms, 
the distorted configuration presents one larger Zn-C bond
(by $\sim$3.5\%) and two shorter bonds ($\sim$5\%)
compared with the bond length (1.99~\AA) of the undistorted
geometry. The distorted configuration is more stable by
160~meV (with SIESTA, this energy difference is reduced to 120~meV using {\sc VASP}). 
This rather small energy difference between the two
configurations might point to the appearance of non-adiabatic electronic effects at room temperature.

The binding energies of the studied
substitutional metal atoms in graphene are
shown in Fig.~\ref{fig:fig2-njp}(d). 
In general, 
the binding energy correlates with 
the carbon-metal bond length, although
the former exhibits a somewhat more complicated
behavior.
The binding energies for transition metals
are in the range of 8-6~eV. 
Subtitutional 
Ti presents the maximum binding energy,
which can be easily understood since for this element
all the metal-carbon bonding
states (Santos et al. 2010b)
are fully occupied. One could expect
a continuous decrease of the binding energy as we
move away from Ti along the transition metal series,
and the
non-bonding 3$d$ 
and the metal-carbon antibonding levels
become progressively populated.
However, the behavior is non-monotonic and the smaller binding
energies among the 3$d$ transition metals
are found for Cr and Mn, 
and
a local maximum
is observed for Co. This complex behavior is 
explained by
the simultaneous energy down-shift
and compression of the 3$d$ shell of the metal impurity
as we increase the atomic number 
(Santos et al. 2010b). 
In summary, 
the binding energies
of the substitutional 3$d$ transition metals 
are
determined by
two competing effects: ({\it i}) as 
the 3$d$ shell becomes occupied and moves to lower energies the hybridization with the
carbon vacancy states near the Fermi energy (E$_F$) is reduced, which decreases the binding energy;
({\it ii}) the transition from Mn to late transition metals is accompanied by a 
reduction of the metal-carbon bond length
by $\sim$0.1~\AA, which increases the carbon-metal interaction and, correspondingly, the binding energy.

The binding
energies for noble metals are considerably smaller than for
transition metals and mirror the reverse behavior of the bond
lengths: 3.69, 1.76 and 2.07~eV, respectively, for Cu, Ag and Au.
The smallest binding energy ($\sim$1~eV) among the metals studied
here is found for Zn, with 
both $s$ and $d$  
electronic shells filled.

\subsection{Spin Moment Formation: 
Hybridization between Carbon Vacancy and 3$d$ Transition Metal Levels}
\label{substitutionals-spins}

\begin{figure}\centering
\includegraphics[width=3.950in]{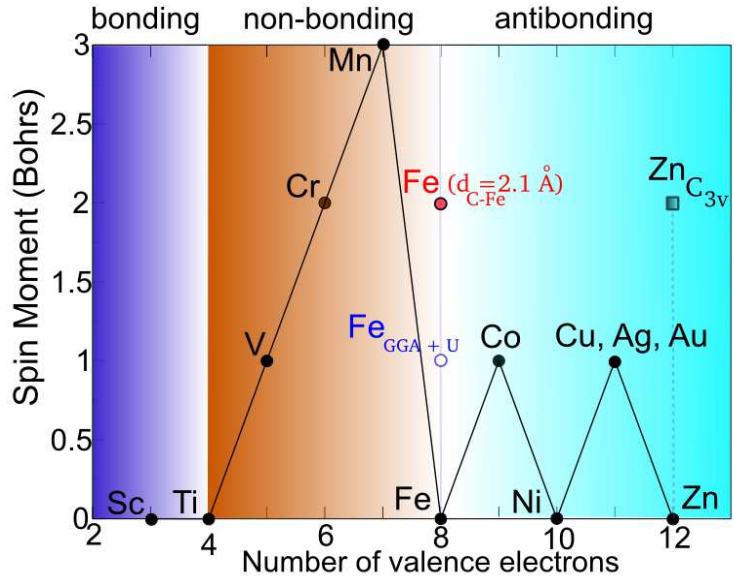}\centering
\caption{\label{fig:fig3-njp}  Spin moment of
substitutional transition and noble metals in graphene
as a function of the number of valence electrons (Slater-Pauling-type
plot).
Black symbols correspond to the
most stable configurations using GGA. Results
are almost identical using {\sc Siesta} and {\sc Vasp} codes. Three
main regimes are found as explained in detail
in the text: ({\it i}) filling of the metal-carbon bonding
states gives rise to the non-magnetic behavior of Ti and Sc;
({\it ii}) non-bonding $d$ states are filled for V, Cr and
Mn giving rise to high spin moments; ({\it iii})
for Fe all non-bonding levels are occupied and
metal-carbon antibonding states start to be filled
giving rise to the observed oscillatory behavior for
Co, Ni, Cu and Zn.
Open and 
red symbols correspond, respectively, to
calculations of Fe using GGA+U and artificially increasing
the height of the metal atom over the graphene layer (see the text).
Symbol marked as Zn$_{{\rm C}_{3v}}$ corresponds to a Zn impurity
in a high-spin symmetric C$_{3v}$ configuration. Adapted from (Santos et al. 2010b).
}
\end{figure}

Our results for the spin moments of substitutional transition and noble metals in graphene are
shown in Fig.~\ref{fig:fig3-njp}~(Santos et al. 2010b). 
Similar results have been found by several 
authors (Krasheninnikov et al. 2009, Huang et al. 2011).
We have developed a simple model that allows to understand 
the behavior of the spin moment, as well as the main features of the electronic
structure, of these impurities~(Santos et al. 2010b). 
Our model is based on the hybridization of the 
3$d$-states of the metal atom 
with the defect levels of a carbon vacancy in graphene. 
In brief, we can distinguish 
three different regimes according to 
the filling of electronic 
levels:

\begin{itemize}
 \item {\bf bonding regime}: all the carbon-metal bonding levels are filled for
Sc and Ti and, correspondingly, 
their spin-moments are zero;

\item {\bf non-bonding regime}: non-bonding 3$d$ levels become
populated for V and Cr  giving rise to a spin moment of, respectively, 1 and 2~$\mu_B$
with a strong localized $d$ character. For Mn one additional electron is added
to the antibonding $d_{z^2}$ level and the spin moment increases to 3~$\mu_B$;

\item {\bf antibonding regime}: finally, for Fe and heavier atoms
all the non-bonding 3$d$ levels are occupied and the spin moment 
oscillates between 0 and 1~$\mu_B$ as the antibonding metal-carbon levels become occupied.
\end{itemize}

The sudden decrease of the spin moment from 3~$\mu_B$ for Mn to
0~$\mu_B$ for Fe is characterized by a transition from a complete
spin-polarization of the non-bonding 3$d$ levels to a full
occupation of those bands. However, this effect depends on the ratio
between the effective electron-electron interaction within the 3$d$
shell and the metal-carbon interaction 
(Santos et al. 2010b). 
If the hybridization with the neighboring atoms is artificially reduced,
for example by increasing the Fe-C distance, Fe impurities develop a
spin moment of 2~$\mu_B$ 
(see the red symbol in Fig.~\ref{fig:fig3-njp}). 
Our results also show that it is 
possible to switch on the spin moment of Fe by changing the
effective electron-electron interaction within the 3$d$ shell. 
These
calculations were performed 
using the so-called GGA+U method. For a large enough
value of U (in the range 2-3~eV), Fe impurities develop a spin
moment of 1~$\mu_B$. It is 
noteworthy
that this behavior is unique to 
Fe: using similar values of U for other impurities does not modify their spin moments.

\begin{table}\centering
\caption{\label{tab:spinmoment}
Mulliken population analysis of the spin moment
in the central
metal impurity (S$_M$) and the carbon nearest neighbors
(S$_C$) for different substitutional impurities in graphene.
S$_{tot}$ is the total spin moment in the supercell.}
\begin{tabular}{@{}lccc}
\hline
     &  S$_M$($\mu_B$)  & S$_C$ ($\mu_B$)  & S$_{tot}$ ($\mu_B$)  \\
\hline
V    & 1.21   & -0.09 & 1.0 \\
Cr     & 2.53 & -0.20 & 2.0 \\
Mn     & 2.91 & -0.10 & 3.0 \\
Co     & 0.44 & 0.06  & 1.0 \\
\hline
Cu     &
      0.24
 & -0.03, 0.31, 0.31
 &     1.0             \\
Ag   &      0.06
     & -0.31, 0.54, 0.54
     &  1.0            \\
Au  &    0.16
    &   -0.28, 0.50, 0.50
    &  1.0      \\
\hline
    Zn$_{{\rm C}_{3v}}$  &    0.23
    &  0.37
    &  2.0    \\
\hline
\end{tabular}
\end{table}

At the level of the GGA 
calculations,
Fe constitutes the border between two different
trends
of the spin moment associated with the substitutional metal impurities in graphene.
For V, Cr and Mn the spin moment is mainly 
due to the polarization of the 3$d$ shell of the
transition metal atoms.  
The strongly localized character of the spin moment for those impurities, 
particularly for V and Cr,  
is corroborated by the Mulliken population analysis shown 
in
Table~\ref{tab:spinmoment}.
For Co, Ni, the noble metals and Zn
the electronic levels close to the
E$_F$ have a 
much 
stronger contribution from 
the carbon 
neighbors.
Thus, for those impurities we can talk about a ``defective graphene"-like magnetism.
Indeed, it is possible to draw an analogy between the electronic structure 
of the late transition, noble metals and Zn substitutional impurities
and that of 
the isolated 
unreconstructed (D$_{3h}$) carbon vacancy 
(Santos et al. 2008, Santos et al. 2010a, Santos et al. 2010b). 
The stronger carbon contribution and delocalization 
in the distribution of the spin moment for Co, the noble metals and Zn 
impurities is evident in Table~\ref{tab:spinmoment}.

\begin{figure}\centering
\includegraphics[width=3.75250in]{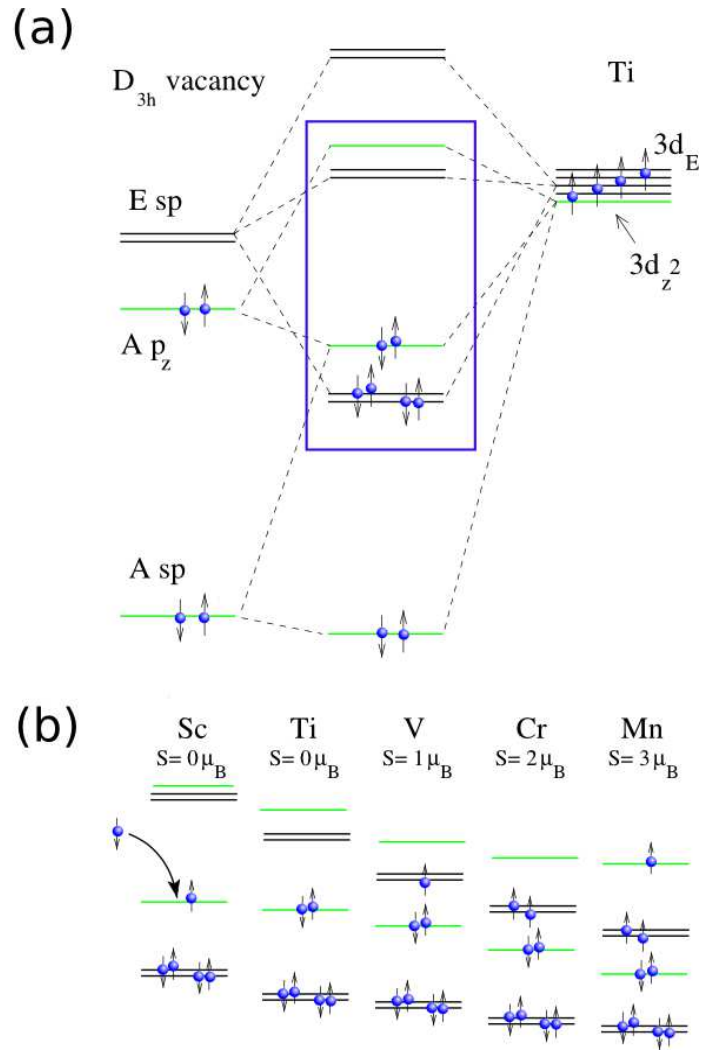}
\caption{\label{fig:fig5-njp} (a) Scheme of the
hybridization between the 3$d$ levels of Ti and the localized
impurity levels of the D$_{3h}$ 
carbon vacancy. Only $d$ levels of Ti are
represented since our calculations show that, at least for
transition metals, the main contribution from $s$ levels appears
well above E$_F$. Levels with A symmetry are represented by gray
(green) lines, while those with E symmetry are marked with black
lines. The region close to E$_F$ is highlighted by a 
square.
(b) Schematic representation of the evolution of the electronic
structure near E$_F$ for several substitutional transition metals in
graphene. The spin moment (S) is also indicated. Substitutional Sc
impurities act as electron acceptors, causing the p-doping of the
graphene layer. Adapted from (Santos et al. 2010b)}
\end{figure}

In the following we present the ``hybridization'' model that
allows to distinguish the three 
regimes of
the spin-moment evolution described before, corresponding 
to the filling of levels of different character.  We have found 
that the electronic structure of 
the substitutional impurities can be easily understood
as a result of the interaction of two 
entities:
(i) the
localized defect
levels 
associated with a symmetric $D_{3h}$ carbon vacancy and, (ii) 
the 3$d$ states of the metal atom, 
taking also into account the 
down shift 
of the 3$d$ shell as the atomic number increases. 
We 
considered
explicitly the 3$d$ states of the metal atom since our calculations show that,
at least
for transition metals, the main contribution from 4$s$ orbitals appears well above E$_F$.

To illustrate the main features of our model  
in Figure~\ref{fig:fig5-njp}~(a) we present a schematic representation  of
the hybridization of the 3$d$ levels of Ti with those of an
unreconstructed $D_{3h}$ carbon vacancy in graphene. 
The interested reader can see (Santos et al. 2010b) 
for an extension of the model for the other metals and technical details. 
The defect levels of the unreconstructed D$_{3h}$ vacancy can be
easily classified according to their $sp$ or $p_z$ character
and whether they transform according to A or E-type
representations. A scheme of 
the different level can be found in Fig.~\ref{fig:fig5-njp}, while the
results
of a DFT calculation are depicted in Fig.~\ref{fig1-co}~(b) (see also
Santos et al 2010b and Amara et al 2007).
Close to the E$_F$ we can find a
fully symmetric A $p_z$ level (thus belonging
to the A$^{\prime\prime}_2$ irreducible representation of
the D$_{3h}$ point group)
and two degenerated
defect levels with E symmetry and $sp$ character
(E$^\prime$ representation).
Approximately 4~eV below E$_F$ we find another defect level
with A $sp$ character (A$^\prime_1$ representation).
Due to the symmetric position of the metal
atom over the vacancy the system has a $C_{3v}$ symmetry and the
electronic levels 
of the substitutional defect
can still be classified according to the $A$ or $E$
irreducible representations of this point group. Of course, metal
and carbon vacancy states 
couple 
only
when they belong to the same
irreducible representation. Thus, occupied $A$ $p_z$ and $A$ $sp$
vacancy levels can only hybridize with the 3$d_{z^2}$ orbitals
($A_1$ representation), while all the other 3$d$ metal orbitals can
only couple to the unoccupied $E$ $sp$ vacancy levels.

With these simple rules in mind and taking into account
the relative energy position of carbon and metal levels, that
changes as we move along the transition metal series,
we can 
understand the electronic structure of substitutional transition metals in graphene
as represented in Fig.~\ref{fig:fig5-njp}~(a) and (b).
Some parameters in the model can be
obtained from 
simple
calculations. For example, a rough estimate of
the position of the 3$d$ shell of the metal atom respect to the
graphene E$_F$ 
is
obtained from the positions of the atomic 
levels.
The relative strengths of the different carbon-metal
hoppings can be estimated from those of the corresponding overlaps.
With 
this
information it is already possible to obtain most
of the features  of the model in Fig.~\ref{fig:fig5-njp}. However,
some uncertainties remain, particularly concerning the relative
position of levels with different symmetry. To solve these
uncertainties the simplest approach is to compare with 
first-principles calculations. 
The details of the model presented in Fig.~\ref{fig:fig5-njp} have been
obtained from a thorough analysis of our calculated band structures (Santos at al. 2010b).
In particular, we have used the projection
of the electronic states into orbitals of different symmetry
as an instrumental tool to classify the levels and to obtain
the rational that finally guided us to the proposed model.
In contrast,
it is interesting to note that some features that derive
from our way to understand the electronic structure of these defects
are very robust and could actually be guessed without direct comparison with 
the calculated band structures. For example, the fact that for
V we start to fill the non-bonding 3$d$ states, 
and
this impurity, as well as Cr and Mn, develops a spin moment,
can be argued from simple symmetry and electron-counting arguments.

According to our model for the substitutional metals, there are three
localized defect levels with $A_1$ character
and three twofold-degenerate levels with $E$ character.
Two of these $E$ levels correspond to
bonding-antibonding $sp$-$d$ pairs, while the third
one corresponds to  3$d$ non-bonding states.
For Sc-Mn the three $A_1$ levels can be pictured as
a low lying bonding level with $A$ $sp$-$d_{z^2}$
character and a bonding-antibonding pair with
$A$ $p_z$-$d_{z^2}$ character.
As
shown in Fig.~\ref{fig:fig5-njp} we have
four metal-vacancy bonding levels (two $A$ and one
doubly-degenerate $E$ levels) that can host up to eight electrons.
For instance,
Ti contributes with four valence electrons, and there are
four electrons associated with the localized carbon-vacancy levels.
Ti has
the bonding states 
completely occupied. Consequently,
Ti presents the highest binding energy among all 3$d$ transition metals and 
has a zero spin moment.
Figure~\ref{fig:fig5-njp}~(b) shows the situation
for other impurities in the series Sc-Mn. Substitutional Sc
impurities have zero spin-moment because they act as electron acceptors.
Note that
all the bonding levels are also filled for Sc, causing a p-doping of the
graphene layer. As already mentioned V, Cr and Mn present an increasing spin moment
due to the filling of the non-bonding levels, while for Fe 
the non-bonding shell is completely filled.

Late transition, noble metals and Zn substitutional impurities
have  the filled
levels coming from an antibonding
interaction between the carbon vacancy and the metal states.
The character and spatial localization of those levels are very similar to those of 
the levels 
of the D$_{3h}$ vacancy close to E$_F$.

Co substitutionals present a singly-degenerate half-occupied defect level at E$_F$.
As we will discuss in more detail in 
the next section,
this level is 
reminiscent
of the state that appears at E$_F$ associated with 
a single carbon vacancy in a $\pi$-tight-binding description
of graphene (Palacios et al. 2008). 
A second electron occupies this level for Ni 
impurities,
and the spin polarization is lost (Santos et al. 2008).

An additional electron is added for noble metal impurities. 
This electron populates a doubly-degenerate level coming
from the antibonding interaction of the 2$sp^{2}$ lobes in the 
nearest carbon neighbors, with the orbitals of d$_{xz}$ and d$_{yz}$
symmetries in the metal impurity.  This state is reminiscent of 
E $sp$ level of the D$_{3h}$
carbon vacancy. The occupation of this two-fold degenerate state
with only one electron explains both, the observed 1~$\mu_B$ spin moment
and the structural distortion of
the noble metal impurities  
(Santos et al. 2010b). As we will see in Section~\ref{strain-tensile}, the
E $sp$ impurity levels also play a crucial role to explain 
the switching on 
of the magnetism of Ni impurities under
mechanical deformations and uniaxial strain.

For Zn two electrons occupy the two-fold degenerate 
E $sp$ level. 
As a consequence, the system suffers a
Jahn-Teller distortion and has a zero spin moment.
However, it is possible to stabilize a symmetric
configuration (Zn$_{{\rm C}_{3v}}$)
with a moment of 2~$\mu_B$ and only
slightly higher in energy 
(Santos et al. 2010b).

\subsection{
Co Substitutionals in Graphene as a Realization of Single $\pi$-vacancies
}
\label{Co-analogy}

\begin{figure}
\includegraphics[width=3.1502400in]{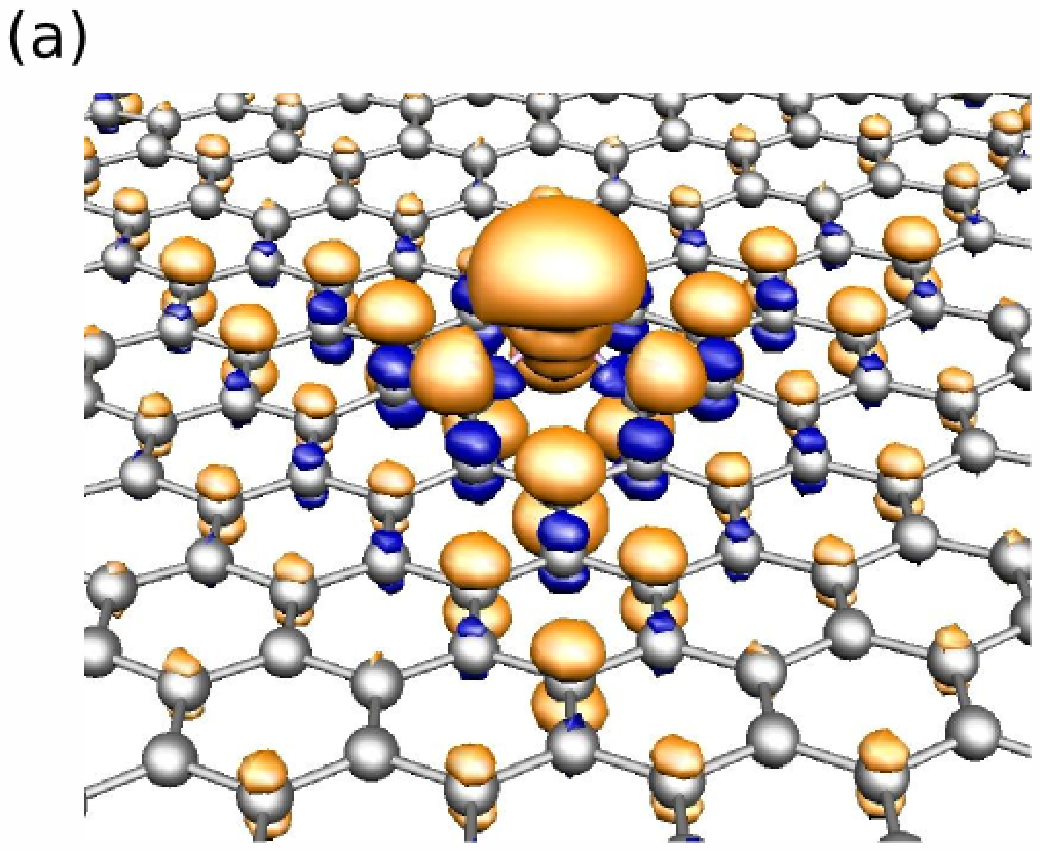}\centering \\
\includegraphics[width=3.627300in]{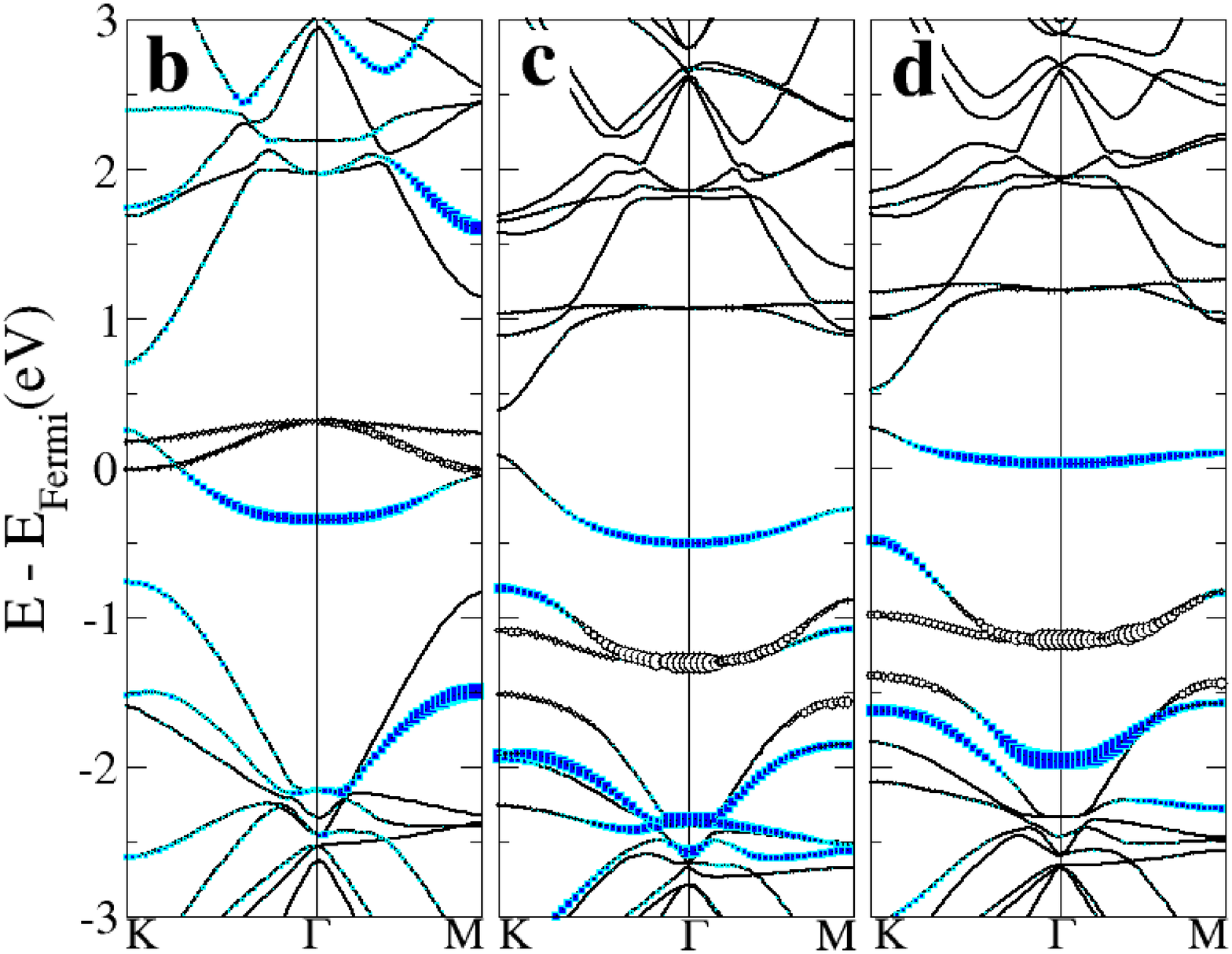}\centering
\caption{(a) Isosurface  of the spin density induced by
a Co$_{sub}$  defect. Positive and negative  spin densities correspond
to   light   and   dark   surfaces  with   isovalues   of   $\pm$0.008
e$^-$/Bohr$^3$, respectively.  Panel (b) presents the spin-unpolarized
band structure of an unreconstructed $D_{3h}$ carbon vacancy. 
Panel (c) and (d) show, respectively,  the band structure of majority 
and  minority spins for a Co$_{sub}$  defect in  a similar  cell.  
The size of filled  symbols in  panel (b) indicate  the contribution 
of the  $p_{z}$  orbitals of  the C  atoms surrounding  the  vacancy, 
whereas  empty  symbols  correspond to  the $sp^{2}$ character. 
In  panels (c) and (d), the filled  and empty circles denote  the 
contribution  of  
hybridized 
Co~$3d_{z^2}$-C~$2p_{z}$  and Co~$3d$-C~$2sp^{2}$ 
characters, respectively. E$_F$ is set to zero. Adapted from (Santos et al. 2010a).}
\label{fig1-co}
\end{figure}

In this section we examine in detail 
the analogy that can be established 
between substitutional Co atoms in graphene (Co$_{sub}$)  
and the simplest theoretical model trying to account for the properties
of a carbon vacancy in graphene. 
The 
electronic structure and magnetic properties of a Co$_{sub}$ impurity 
are analogous to those 
of
a vacancy in a simple $\pi$-tight-binding description of
graphene.
This toy model system, the  $\pi-$vacancy, has been extensively studied 
in the graphene literature due to its very interesting magnetic properties directly
related to the bipartite character of the graphene network 
(Castro Neto et al. 2009, Palacios et al. 2008). 

We begin by looking at the spin
density 
of
the Co$_{sub}$ impurity as shown in 
Figure~\ref{fig1-co}~(a). The  spin polarization 
induced in the carbon 
atoms has a $p_z$-like  shape and decays slowly as we move away from the impurity.  
The sign  of the  spin  polarization 
follows  the  bipartite character of graphene: the polarization aligns 
parallel (antiparallel) to the spin moment located in the Co impurity for 
carbon  atoms in  the opposite (same)  sublattice. The  value of the total  spin  moment is
1.0~$\mu_{B}$ per defect. Using Mulliken  population analysis, 
the moment on the Co atom has
a value of 0.44~$\mu_{B}$; 
the three  first carbon neighbors have 0.18~$\mu_{B}$; and 
there are
0.38~$\mu_{B}$ delocalized in the rest  of the layer.
Therefore, the total spin moment 
has contribution from 
both Co and carbon orbitals. 

To understand the origin of  this spin polarization, we now analyze in
detail the band structure.  Figures~\ref{fig1-co} (c) and
(d)  present the  results  for  a Co$_{sub}$  defect  in a  4$\times$4
graphene supercell.  Similar results are obtained  using a 8$\times$8
cell. For  comparison, panel  (b) shows the  spin-compensated band
structure of  a single unreconstructed $D_{3h}$ carbon vacancy.
For  the  $D_{3h}$  vacancy,  
there are
three
defect  states 
in a  range  of 
$\sim$0.7~eV  around E$_F$.   
Two  states  
appear above  E$_F$
at 0.3~eV at $\Gamma$ 
and
have  a large 
contribution from  the $sp^{2}$ lobes  of the C
atoms surrounding the vacancy. 
These levels correspond to the two degenerate
E~$sp$ states appearing
in Figure~\ref{fig:fig5-njp}.  Another state at 0.35 eV below E$_{F}$ 
shows  a predominant  $p_z$ contribution
and corresponds
to the A $p_z$ level in Figure~\ref{fig:fig5-njp}.  
This last level 
represents
the defect state 
that appears at E$_{F}$ for a
vacancy using a 
$\pi$-tight-binding description.
For a  Co$_{sub}$, the  defect states of the vacancy described
above hybridize with the Co $3d$  states. 
The two $2sp^{2}$ defect
bands,  now an  antibonding combination  of Co~$3d$  and  the original
C~$2sp^{2}$   vacancy   levels,  are   pushed   at  higher   energies,
$\sim$1.0~eV  above  E$_{F}$ (see  Fig.~\ref{fig1-co}~(c)  and (d)).  The
singly  occupied $p_{z}$  state,  now hybridized  mainly  with the  Co 
$3d_{z^2}$ orbital,  remains at E$_{F}$  and becomes almost
fully  spin-polarized. The  Co$_{sub}$  impurity  becomes thus 
analogous to a vacancy in the $\pi$-tight binding model of 
graphene.

This analogy is a powerful one, since it brings our results for the
magnetism of Co$_{sub}$ impurities
into contact with Lieb's theorem for a half-filled
Hubbard model (Lieb 1989), where the spin polarization is an
intrinsic property of the defective bipartite lattice. 
Applying this theorem and our analogy, we can expect 
that the total spin of an array of Co$_{sub}$
impurities can be described according to the simple rule
S=$|N_{A}-N_{B}|$, where $N_{A}$ and $N_{B}$ indicate the number
of Co substitutions in A and B sublattices, respectively. In
Section~\ref{substitutionals-coupling} we will show results from
first-principles calculations that confirm this behavior.
However,
Lieb's theorem is global,
in the sense that it refers to the total 
spin moment of the system, and 
does not enter into the local description of the magnetic interactions. 
This will be described in more detail in Section~\ref{substitutionals-coupling},
where we will compute
the exchange couplings between Co$_{sub}$ defects. 

Other realistic defects, besides Co$_{sub}$ impurities, can also be mapped
onto the simple $\pi$-vacancy model. 
In 
Section~\ref{covalent-spin_formationNT}
we will see that complex adsorbates chemisorbed on carbon nanotubes and graphene
generate a spin polarization. The magnetism due to such 
a
covalent functionalization 
displays a behavior similar to that of the $\pi$-vacancies.
Some concepts already used here will be again invoked to explain the 
main features of the magnetism associated with these defects, leading to a universal   
magnetic behavior independent of the particular adsorbate.

\section{
Tuning the Magnetism of Substitutional Metals in Graphene with Strain}
\label{strain}

In the previous section, 
we have considered in detail the 
formation of local spin-moments
induced by a particular class of defects in graphene, substitutional transition metals. 
Although this is an important subject,  
other aspects are also crucial to understand and control the magnetism 
associated with this kind of doping. For example, one needs to explore 
the characteristics of
the couplings between local moments, as well as the possibility to 
control such couplings, and the size of the local moments, using external
parameters. This kind of  knowledge 
is instrumental 
in
possible applications 
in spintronics and quantum information devices. 
The subject of the magnetic couplings will be postponed until Section~\ref{couplings}.
In this Section we analyze how the structural, 
electronic and magnetic properties of substitutional defects 
in carbon nanostructures can be controlled using 
strain. 
We focus on Ni substitutionals and conclude that 
externally applied strain can provide 
a
unique tool to tune
the magnetism of Ni-doped graphene.

Strain provides a
frequently used 
strategy to modify the properties of materials. 
For example, strain is intentionally applied to 
improve mobility in modern microelectronic devices. This so-called strain engineering has 
taken a key position over years. Recently, strain effects have also 
been proposed as a route to control the electronic properties of 
pristine 
graphene, which 
had a deep impact 
on the physics of this material~(Guinea et al. 2009, Pereira et al. 2009).

Here,
we show that the application of uniaxial strain 
can be used
to switch on the magnetism of graphene doped with Ni substitutional 
impurities (Ni$_{sub}$) (Santos et al. 2012a). 
Whereas Ni$_{sub}$ 
defects
are non-magnetic in flat graphene, 
we find that 
their spin moment changes from zero 
when no strain in applied, up to 1.9 $\mu_{B}$ at $\sim$7.0\% strain.  
This strong variation 
stems from 
the modifications of the local structure of 
the defect, which  
cause changes in the electronic structure of 
the defect that
can be related to those of the unreconstructed 
carbon vacancy in graphene under strain. 
The similarities between the electronic structure of the 
D$_{3h}$ vacancy and that of Ni$_{sub}$
were already stressed in the previous section.

We also show that substitutional metallic impurities in carbon nanotubes
display a different magnetic behavior from that observed in flat graphene.
Using Ni$_{sub}$ dopants as an example we demonstrate that 
the intrinsic curvature
of the carbon layer
in the SWCNTs can be used to switch on the magnetism of Ni substitutionals (Santos et al. 2008).  
The defect electronic structure is modified by curvature 
in a similar way 
as by uniaxial strain.
In addition, we find
a strong dependence 
of the spin moment on the impurity distribution, tube metallicity 
and diameter of the nanotube.

\subsection{Switching On the Magnetism of Ni-Doped Graphene with Uniaxial Strain}
\label{strain-tensile}

In this subsection,
we study 
the electronic 
structure of Ni$_{sub}$ defects in graphene
under 
uniaxial
strain.
According to the analysis presented in Section~\ref{substitutionals-spins},
at 
a
zero strain 
Ni$_{sub}$ defects are non-magnetic in flat graphene~(Santos et al. 2008, Santos et al. 2010b).
However, we find that 
under 
moderate uniaxial strain these impurities develop
a non-zero spin moment, whose size increases with that of the applied strain. 
This magnetoelastic effect 
might be utilized
to design a strain-tunable spin device based on defective graphene.

\begin{figure}
\includegraphics[width=5.00in]{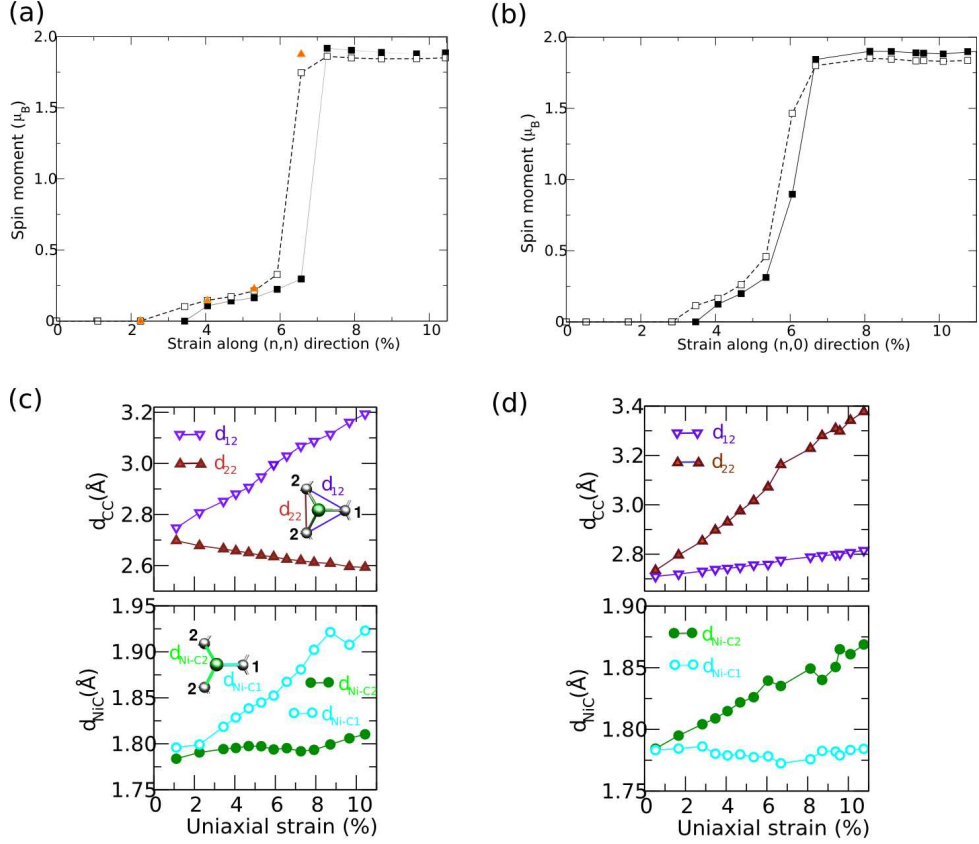} 
\caption{Spin moment as a function of the applied strain along 
(a) the $(n,n)$ and (b) the $(n,0)$ directions. In panels (a) and (b), filled squares indicate results 
obtained using geometries from a non-spin polarized calculation using a DZ basis. The spin 
moment and electronic structure is calculated using a DZP basis using such geometry.
Open squares indicate a similar calculation, but the geometries have been obtained from a spin-polarized 
calculation. The triangles represent full calculations (geometry and spin moment)
with DZP basis set. 
Panels (c) and (d) show the 
results for the structural parameters
as a function of the applied strain along $(n,n)$ and $(n,0)$, respectively. Bond lengths between the different C atoms are denoted d$_{12}$ and d$_{22}$, while the bond lengths 
between Ni and C atoms are d$_{Ni-C1}$ and d$_{Ni-C2}$. The structural information
was calculated using a DZ basis.
Adapted from (Santos et al. 2012a). 
}
\label{fig2-ni-strain}
\end{figure}

Figures \ref{fig2-ni-strain}~(a)-(b) show the spin moment of a Ni$_{sub}$ defect 
as a function of the applied strain along the $(n,n)$ and $(n,0)$ directions, respectively. 
The curves with filled squares show simulations using geometries from a non-spin polarized calculation 
with a DZ basis set 
(see references Soler et al. 2002 and Artacho et al. 1999 for a description 
of the different basis sets). The spin moment and electronic structure are always 
calculated using a more complete DZP basis.
The open squares indicate systems that were calculated using the previous procedure, i.e. a DZ basis, but the 
geometries have been obtained from spin-polarized calculations. The triangles display calculations
with DZP basis set for both geometry and spin moment. 
At zero strain the Ni$_{sub}$ defect is non-magnetic.
As the uniaxial tension is applied, the system starts to deform. 
At $\sim$3.5\% strain,
the system becomes magnetic with a 
spin moment that evolves nearly linearly with the uniaxial strain up to  
values of $\sim 0.30 - 0.40$~$\mu_{B}$ at $\sim$6.0\%.  
The 
magnetism
of the system using different basis set is very similar.
At $\sim$6.8\% the spin moment increases sharply from $\sim$0.40~$\mu_{B}$ to $\sim$1.9~$\mu_{B}$. 
This transition 
takes place
for both directions, although it is somewhat more abrupt 
along the $(n,n)$ direction (Figure \ref{fig2-ni-strain}(a)) where no intermediate steps are observed. 
Thus, 
the magnetic properties depend on 
the local defect geometry and,  
to a much lesser extent, on
the defect orientation relative to the applied 
strain.
Panels (c) and (d)
in Figure \ref{fig2-ni-strain} present  
the local defect geometry. When the strain is applied, the triangle formed 
by the three C neighboring atoms to the Ni impurity deforms. 
The
C-C distances along the strain direction 
increase, whereas distances along the perpendicular direction decrease in response to such elongation.
The distance of the Ni atom to the first carbon neighbors also increases, but this bond length changes 
for the studied strains are less than $\sim$5.0\% (
averaged over both strain directions) in comparison with 
$\sim$20.0\% for the C-C distances.
The analysis of distances suggests that 
the carbon neighbors 
and the central Ni impurity 
interact strongly, 
which is also
reflected in the high stability of the defect
with a binding energy 
$\sim$7.9~V 
to the carbon 
vacancy.

\begin{figure}
\includegraphics[width=3.200in]{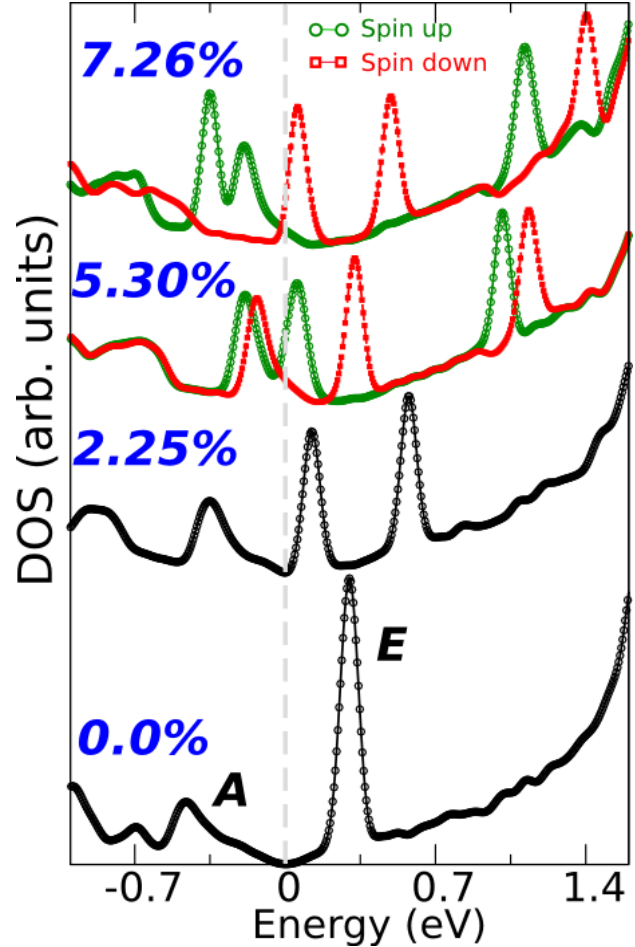} \centering
\caption{
Density of states (DOS) of the Ni$_{sub}$ 
defect under 0.0\%, 2.25\%, 5.30\% and 7.26\% strain applied along the $(n,n)$ 
direction. Symbols $A$ and $E$ indicate the character and symmetries of 
the defect states, with large weight of Ni hybridized with C states. 
$A$ corresponds to Ni $3d_{z^2}-$C $2p_{z}$, and 
$E$ represents Ni $3d_{xz}, 3d_{yz}-$C $2sp$. 
At 5.30\% and 7.26 \% strains, the open squares (green curve) represent the spin up channel 
and filled squares (red curve) the spin down. For clarity, the curves 
have been shifted. The Fermi energy is marked by the dashed (gray) 
line and is set to zero. Adapted from (Santos et al. 2012a).}
\label{fig3-dos-ni-strain}
\end{figure}

In order to understand 
the
magnetic moment in Ni$_{sub}$ defects,
the density of states (DOS) around E$_F$ under strains
of 0.0\%, 2.2\%, 5.3\% and 7.2\% are shown in Figure \ref{fig3-dos-ni-strain}. 
The strain is along the 
$(n,n)$ direction although the qualitative behavior is similar 
to
other directions. 
Several 
defect levels 
around E$_F$ have Ni and C mixed character.
As pointed out before (Section~\ref{substitutionals}), 
the metal atom over the vacancy
has a $C_{3v}$ 
symmetry at 
a
zero 
strain, and
the electronic levels are 
classified according 
to the $A$ or $E$ irreducible representations of this point group. 
Essentially, these three defect states and their evolution as a function 
of the applied strain determine all the observed physics.  

One of them with $A$ character is occupied and appears around $\sim$0.50~eV 
below E$_F$ at 
a
zero strain. This level comes from a fully symmetric 
linear combination of the $2p_z$ orbitals (z-axis normal to the layer) 
of the  nearest C neighbors interacting with the $3d_{z^2}$ orbital
of Ni. The other twofold-degenerate levels with $E$ character, coming from the hybridization 
of the in-plane $sp^2$ lobes of the carbon neighbors with the 
Ni $3d_{xz}$ and $3d_{yz}$ orbitals, appear at 0.50~eV above E$_F$ at 
a
zero strain. 
Because this electronic structure has 
the Ni $3d$ states far from E$_F$
and no flat bands crossing E$_F$, the
spin moment of the Ni$_{sub}$ impurity in graphene is zero. 
Interestingly, these three levels that appear close
to E$_F$ in Figure \ref{fig3-dos-ni-strain}
are reminiscent of those found 
for the unreconstructed carbon vacancy in graphene as we have already seen in Section \ref{substitutionals}.

The energy position of the three levels shifts as a function of the 
applied strain. When the strain is applied, the degeneracy between 
Ni $3d_{xz}-$ C $2sp$ and Ni $3d_{yz}-$ C $2sp$ states is removed 
and a gradual shift towards E$_F$ of one of them is observed.
This level becomes partially populated, and the system starts to develop 
a spin moment. The Ni $3d_{z^2} -$C $2p_{z}$ state also changes 
its position approaching $E_F$, although 
for small values of the strain this level does not contribute to the
observed magnetization. However, 
around a 7\% strain
both the Ni $3d_{z^2}-$ C $2p_{z}$ and the Ni $3d_{xz,yz}-$ C $2sp$ levels become 
fully polarized and the system 
develops 
a moment close to 2.00 $\mu_{B}$. 

Figure \ref{fig3-dos-ni-strain} also shows the resulting 
spin-polarized DOS at 5.3\% and 7.2\% strain (upper part of the panel).  
The exchange splittings of the $3d_{xz}$ and $3d_{yz}$ levels are, respectively, 
$\sim$0.29~eV and $\sim$0.13~eV at 5.3\%, 
increasing with the applied strain and the associated spin moment.
The
energy gain with 
respect to the 
spin-compensated solutions 
develops
from 13.9 meV at 5.3\%  to 184.1 meV at 7.26\%. 
Thus, a
moderate variation of the strain applied 
to
the graphene layer 
changes
the 
spin 
state
and enhances the stability of the defect-induced moment. 
According to these results, 
if 
it
is possible to control the strain applied 
to
the graphene layer,
as shown in recent experiments (Mohiuddin et al. 2009, Kim et al. 2009), 
the magnetism 
of Ni-doped graphene 
could be turned {\it on} and {\it off} 
at will, 
like switches used in {\it magnetoelastic devices}, 
however, with no applied magnetic field. This suggests a sensitive and effective way to control 
the magnetic properties of graphene, which is interesting for its possible applications 
in nanoscale devices (Santos et al. 2012a).

\begin{figure}
\includegraphics[width=5.00in]{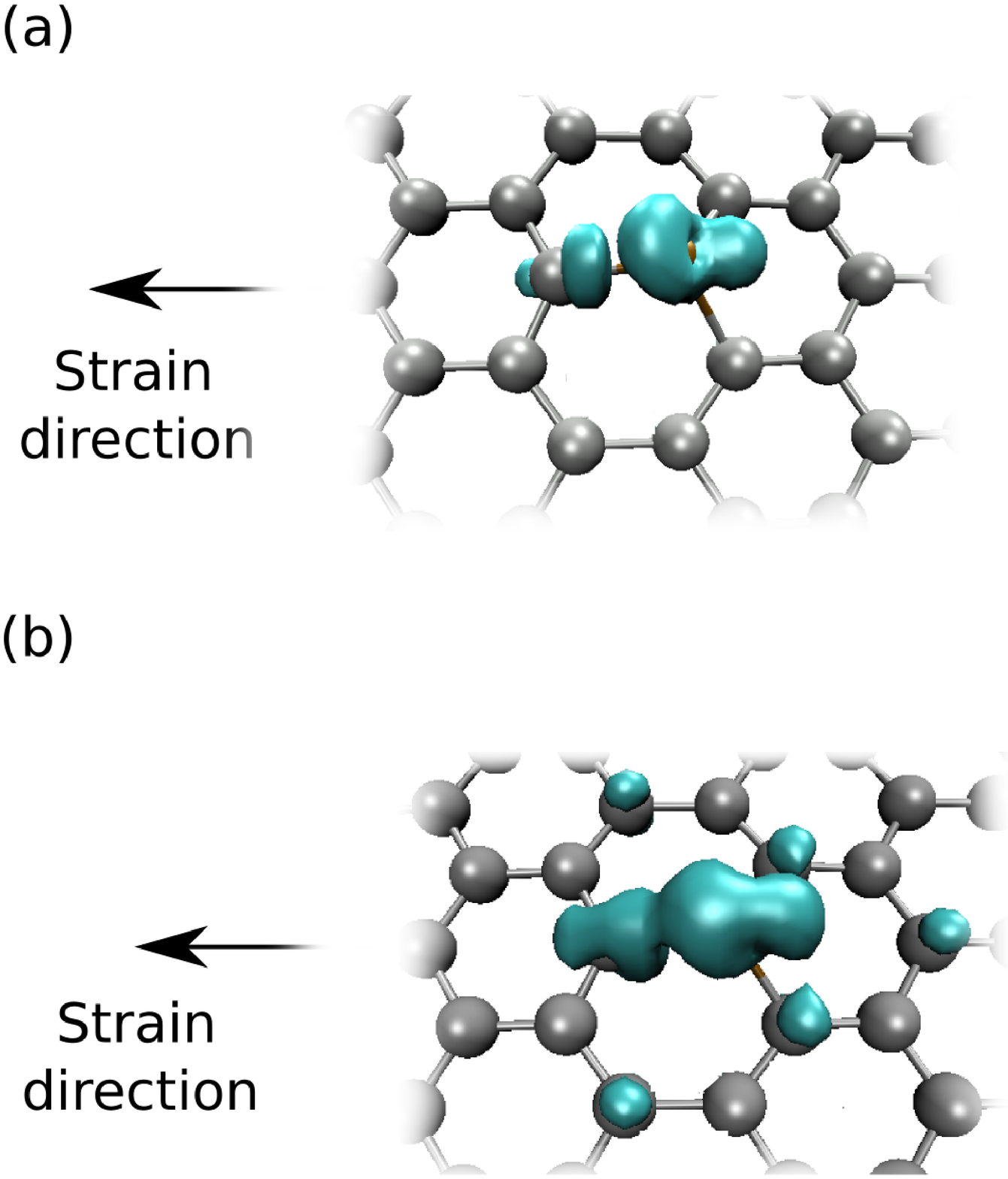} 
\caption{(a)-(b) Spin densities for Ni$_{sub}$ defects at 
strains of 5.30\% and 7.26\% along the $(n,n)$ direction. The strain direction
is marked by the arrows in both panels. The isovalue cutoff at (a) and (b) panels 
is $\pm$0.035 and $\pm$0.060 e$^-$/Bohr$^3$. Adapted from (Santos et al. 2012a).
}
\label{fig4-ni-strain}
\end{figure}

Figure \ref{fig4-ni-strain}(a)-(b) 
shows
the spin magnetization patterns induced by 
the presence of a Ni$_{sub}$ defect under two different magnitudes of uniaxial strain applied along the $(n,n)$ direction. 
The spin polarization induced in the neighboring carbon atoms
has  
shape and orbital 
contributions that depend
sensitively on 
strain.
At 5.30\% the spin density is mainly 
localized at the Ni impurity and at the C atom 
bonded to Ni along the strain direction. 
The anti-bonding character of the $E$ defect state that 
originates the magnetization is 
clear (see the node 
in
the bond direction).
The spin density at this C atom shows
a $2sp$-like shape to 
be contrasted with that at 7.26\% strain, in which apart from
the $2sp$-like shape, a $2p_z$ component is clearly observed. At this
larger strain
farther
neighboring-carbon atoms also contribute
to the spin density 
with mainly $2p_z$ character. 
This additional contribution to the 
spin polarization pattern 
corresponds
to 
the Ni $3d_{z^2}-$C $2p_{z}$-defect state at E$_F$ for strains above $\sim$7\%, 
as 
explained in the previous section using the DOS.

\begin{figure}
\includegraphics[width=3.500in]{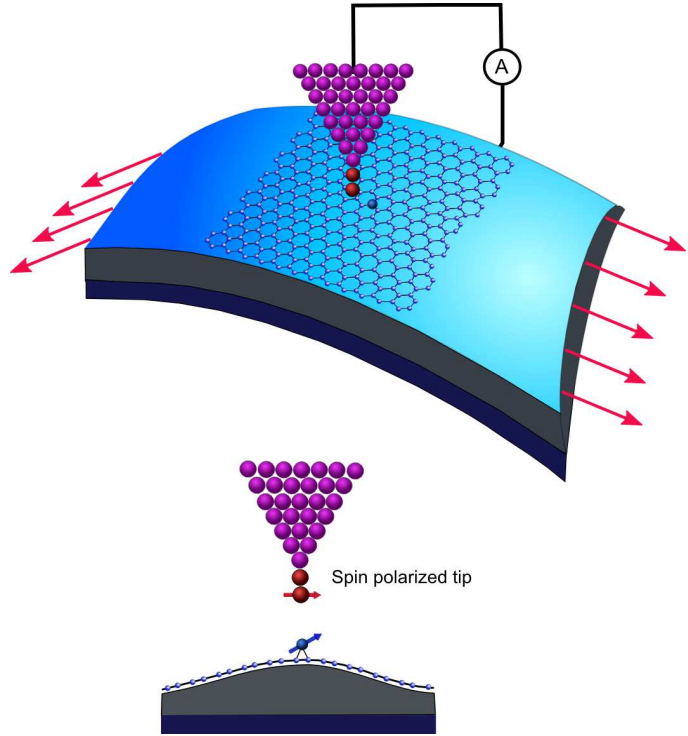} \centering
\caption{Experimental setup 
that could be utilized to measure the effect of strain on 
the magnetic properties of Ni-doped graphene. The layer is deposited on a 
stretchable substrate which keeps a large 
length-to-width 
ratio 
in order to obtain a uniform tensile strain 
in
the graphene film. 
Spectroscopy measurements using 
scanning tunnelling microscope (STM) would allow to 
identify the shift of the different defect levels. If the magnetic anisotropy is
large enough or there is an external magnetic field, it could be also 
possible to measure the 
presence and orientation of a magnetic moment 
at the defect site
using a spin-polarized tip.
}
\label{fig-exp-ni-strain}
\end{figure}

Figure \ref{fig-exp-ni-strain} shows 
a possible experimental setup that could be used to test our predictions.
This is similar to a mechanically controlled break junction setup with an
elastic substrate (Mohiuddin et al. 2009, Kim et al. 2009). 
Graphene is deposited 
in
the center of such 
a
substrate in order to obtain a uniform strain. 
Bending or stretching the substrate causes an expansion of the surface, and the deposited graphene will
follow this deformation. In principle, the modifications on the electronic structure can be detected using 
a scanning tunneling microscope (STM) since the defect levels that are involved are localized around the Fermi energy.  
For example, 
Ugeda {\it et al.} were able to measure using STM
the energy position and spatial localization of the defect levels associated with a carbon
vacancy in the surface of graphite (Ugeda et al. 2010).
If the magnetic anisotropy of the defect is high enough, at sufficiently low temperatures,
a preferential orientation of the moment would be stabilized and, 
in principle, a STM with 
a spin polarized tip (Spin-STM) could allow to monitor the evolution of the magnetic 
properties of the Ni-doped graphene with strain. Instead, an external magnetic field may be used to align the 
magnetic moments of the defects and define the hard/easy 
axis of the system. 
It is 
noteworthy
that the break junction-like setup has already been 
successfully used (Standley et al. 2008).

\subsection{Ni substitutionals in Carbon Nanotubes: Curvature Induced Magnetism}
\label{strain-NiNT}

\begin{figure}
\includegraphics[width=11.50cm]{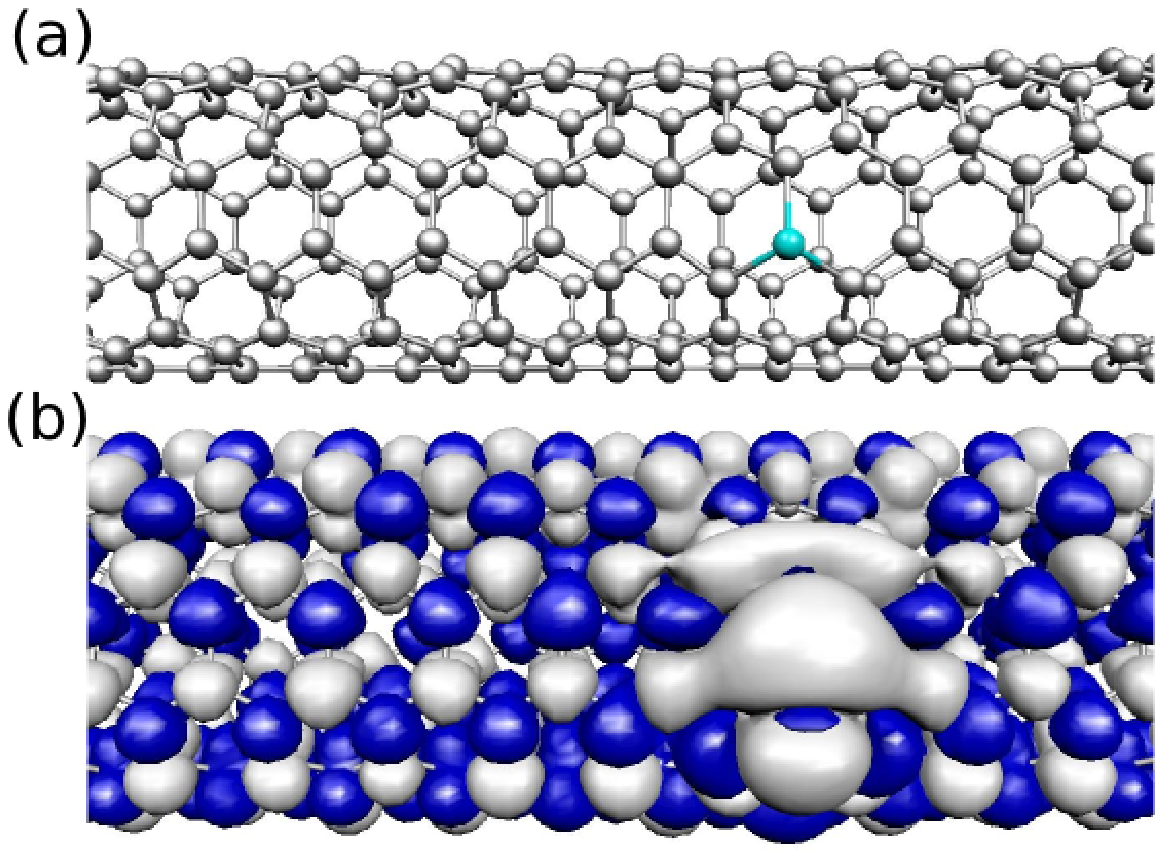}\centering
\caption{  
(a) Relaxed geometry of a substitutional Ni (Ni$_{sub}$) impurity
in a (5,5) SWCNT, and (b) isosurface 
($\pm$0.002~e$^-$/Bohr$^3$)
of the magnetization density with light (gray) and dark (blue) surfaces corresponding,
respectively, to majority and minority spin. Adapted from (Santos et al. 2008).
}
\label{fig:fig1-Ni-tubes}
\end{figure}

Although Ni$_{sub}$ impurities are
non-magnetic in flat graphene, 
their magnetic moment can be switched on by applying curvature
to the structure. To understand why, we will begin 
looking at the equilibrium structure of Ni$_{sub}$ for
a (5,5) SWCNT. 
The Ni atom is
displaced $\sim$0.9~\AA\ from the carbon plane. Although 
both outward and inward displacements were stabilized,
the outward configuration is 
more stable. 
The calculated Ni-C distances (d$_{Ni-C}$)
are in the range 1.77-1.85~\AA\ in good
agreement with experiment~(Ushiro et al. 2006, Banhart et al. 2000).
Armchair tubes exhibit two slightly shorter and one larger values of d$_{Ni-C}$, 
the opposite happens for 
zigzag
tubes, whereas for
graphene we obtain a  threefold symmetric structure with 
d$_{Ni-C}$=1.78~\AA. Ni adsorption inhibits the reconstruction~(Amara et al. 2007)
of the carbon vacancy. Furthermore, we have checked that 
for a vacancy in graphene, 
a symmetric structure is obtained 
after Ni addition
even when starting from a relaxed vacancy. 

Figure~\ref{fig:fig1-Ni-tubes}~(b) shows the magnetization
density profile for a Ni$_{sub}$ defect in 
a (5,5) metallic nanotube at large dilution (0.3~\% Ni concentration).
The total spin moment 
is 0.5~$\mu_B$. 
The magnetization 
is on 
the Ni atom and its 
C neighbors. However, it also extents considerably
along the tube, particularly in the direction perpendicular to the tube axis. 
This 
profile
indicates 
that
the spin polarization 
follows
some of the delocalized
electronic states in the metallic nanotube. Indeed, as we clarify below, the 
magnetism in substitutionally Ni-doped SWCNTs only appears associated with the 
curvature and the metallicity of the host structure.

\begin{figure}
\includegraphics[width=11.50cm]{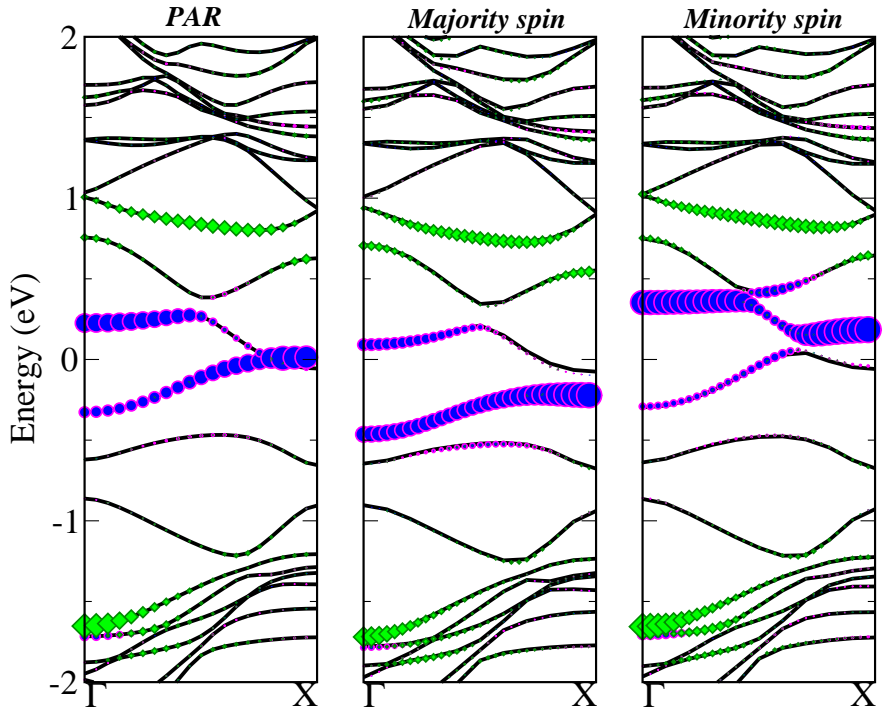}\centering
\caption{
Band structure of a (5,5) nanotube containing
a Ni$_{sub}$ impurity in 
four
unit cells for (left panel) a paramagnetic calculation (PAR),
and for (middle panel) majority and (right panel) minority spins. 
Circles and diamonds correspond respectively to the
amount of Ni 3$d_{yz}$ and 3$d_{xz}$ character.
X-axis is parallel 
to 
the tube axis and y-axis
is tangential. Adapted from (Santos et al. 2008). 
}
\label{fig:fig3-Ni-tubes}
\end{figure}

The basic picture described in Section \ref{substitutionals} is still valid 
for the electronic structure of the Ni$_{sub}$ impurity in SWCNTs.
However, the modifications that appear due to 
the curvature of the carbon layer are responsible 
for the appearance of a magnetic moment.
Figure~\ref{fig:fig3-Ni-tubes} (a) shows the band structure of a paramagnetic
calculation of a (5,5) SWCNT with a Ni$_{sub}$ impurity every four 
unit cells. 
Comparing the 
band
structure in Figure~\ref{fig:fig3-Ni-tubes} (a)
with the electronic structure of the Ni$_{sub}$ impurity in 
flat geographer (lower curve in Figure~\ref{fig3-dos-ni-strain}),
we 
appreciate the effects of curvature.
The degeneracy between d$_{xz}$ and d$_{yz}$ states is removed 
(x-axis taken along the tube axis and y-axis along the tangential 
direction at the Ni site). 

The d$_{yz}$ contribution is stabilized by several tenths of eV 
and a  quite flat band with strong d$_{yz}$ character is
found {\it pinned} at E$_{F}$ close to the Brillouin-zone boundary.
Under these conditions,
the spin-compensated solution 
becomes unstable and a magnetic moment of 0.48~$\mu_B$ is developed. 
Figures \ref{fig:fig3-Ni-tubes}~(b) and (c) show, respectively, 
the band structure for majority and minority spins. The exchange splitting 
of the d$_{yz}$ level is $\sim$0.4~eV and the energy gain with respect to the paramagnetic solution is 32~meV.

\begin{figure}
\includegraphics[width=9.0cm]{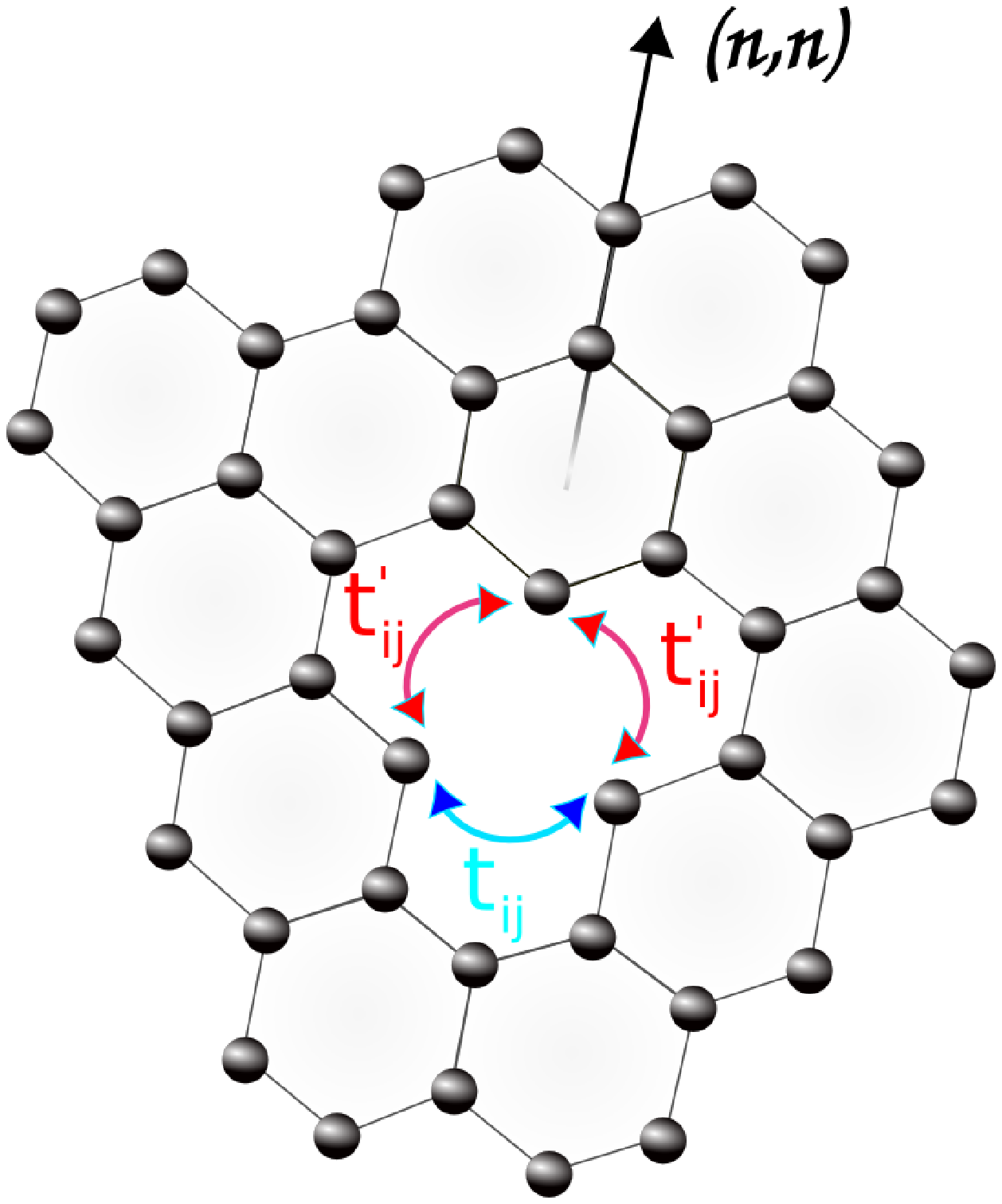}\vspace{0.70cm} \centering
\includegraphics[width=12.0cm]{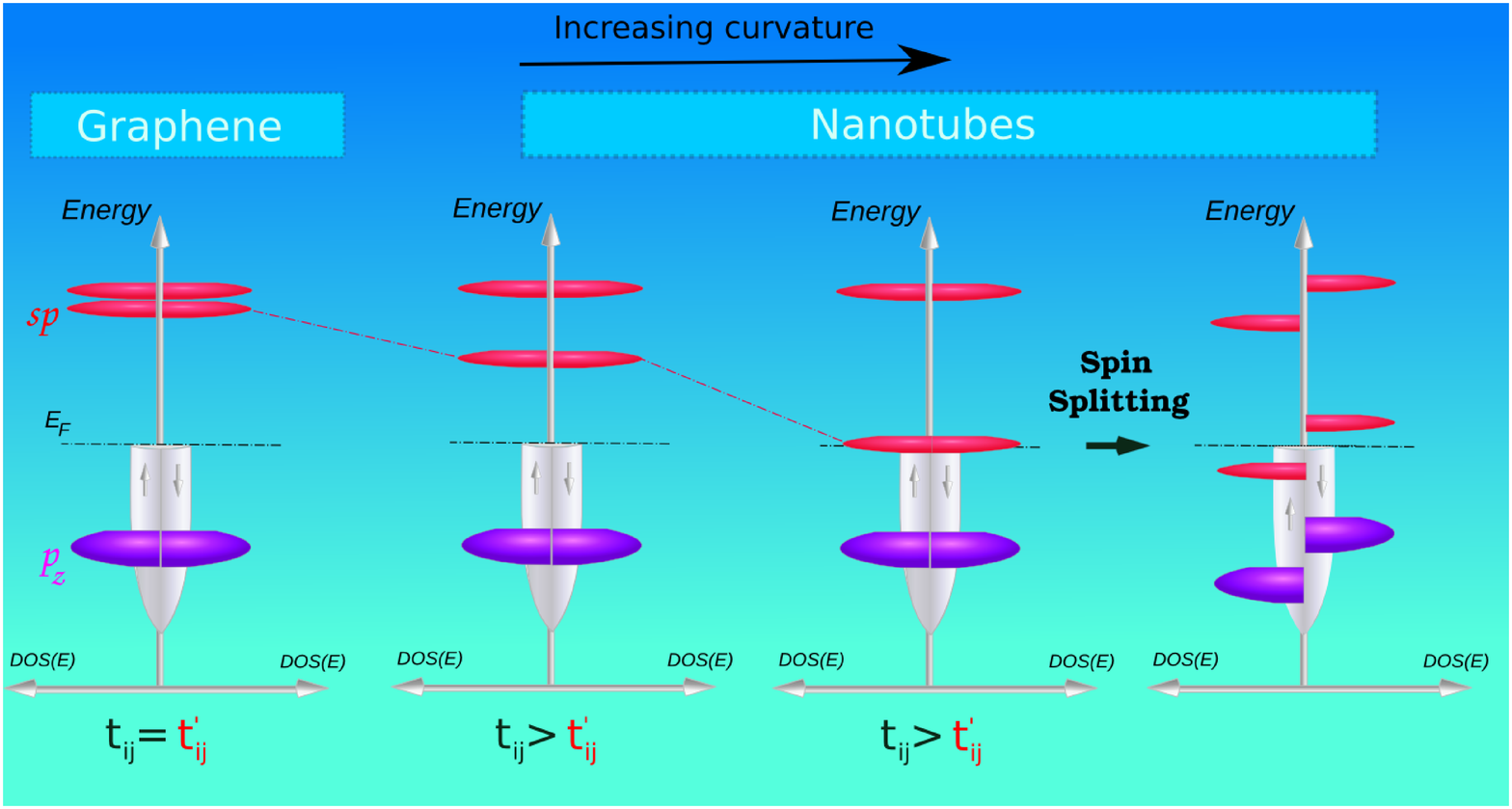}\centering
\caption{
Effect of curvature (anisotropic strain)
on Ni$_{sub}$ in (n,n) tubes. 
Upper panel: Illustration of the 
dominant 
hoppings  at the defect site in graphene. 
The equivalence between the electronic structure of a Ni$_{sub}$ impurity
and a carbon vacancy is stressed here.
The carbon sheet is rolled around the $(n,0)$ direction in order to form the
armchair tubes. 
Lower panel: Scheme of the main Ni$_{sub}$ impurity energy levels 
as a function 
of curvature. 
One of the impurity levels with antibonding C 2$sp$-Ni 3$d$ character
is shifted downwards and, for large enough curvatures, becomes partially populated and spin-polarized. Adapted from (Santos et al. 2008).
}
\label{fig:fig4-Ni-tubes}
\end{figure}

In general, whenever a flat
band with appreciable Ni 3$d$ character
becomes partially filled we can expect the appearance of a spin moment. 
The population of such an impurity level 
will take place at the expense of the simultaneous depopulation of some of 
the delocalized carbon $2p_z$ levels 
within
the host structure.
For this reason,
the development of a 
spin moment is more likely for 
Ni$_{sub}$ impurities in metallic structures like the armchair tubes.
The crucial role of the host states also explains the delocalized character of the magnetization
density depicted in  Fig.~\ref{fig:fig1-Ni-tubes}~(b). However, it is important 
to stress that 
the
spin moment associated with a Ni$_{sub}$ impurity in SWCNTs 
forms
driven by the local curvature of the carbon layer, 
because the energy position of one of the impurity levels 
shifts downwards until it crosses E$_{F}$. 
A schematic representation of this phenomenon is
shown in Fig.~\ref{fig:fig4-Ni-tubes} where we also emphasize the similarities between
the levels of the Ni$_{sub}$ defect and those of the unreconstructed carbon vacancy.
Notice the similarities with the effects of 
uniaxial strain described in the previous Section. 
At large tube diameters,
the limit of flat graphene with zero 
spin moment (see Section \ref{substitutionals})
is recovered.

For semiconducting 
tubes,
the situation is somewhat different. 
The $d_{xz}$ and $d_{yz}$ -like levels 
remain unoccupied unless their energies are shifted by a larger amount that pushes one of them 
below the top of the valence band. 
If
the tube has a large enough gap,
the spin moment is
zero irrespective of the tube diameter. We have explicitly checked that a zero
spin moment is obtained for (8,0) and (10,0)  semiconducting tubes for Ni concentrations
ranging from 1.5\% to 0.5\%. The different magnetic behavior of Ni$_{sub}$ impurities
 depending on the 
metallic and semiconducting character of the host structure provides a 
route to experimentally identify metallic armchair tubes.

\section{Magnetic Coupling Between Impurities}
\label{couplings}

In previous Sections, we have considered the formation
of local moments associated with defects in carbon nanostructures,
as well as the use of mechanical deformations to tune the sizes
of such local moments.  Here, we present calculations of the exchange couplings
between the local moments in neighboring defects. This is a necessary step 
to elucidate whether it is possible to induce magnetic order
in these materials, which is
crucial 
in
the application of carbon-based nanostructures in spintronics.
We focus on 
defects that can be mapped onto the simple model
provided by the fictitious $\pi$-vacancy. According to the results
presented in Section~\ref{Co-analogy}, 
Co substitutional impurities belong to this class of  defects.
In this Section we present
another type of impurities that behave according to the same analogy: 
molecules attached to graphene and carbon nanotubes through weakly-polar covalent bonds. 

A $\pi$-vacancy corresponds to a missing $p_z$ orbital in a graphene plane
described 
using a $\pi$-tight-binding model. The magnetic properties of 
the $\pi$-vacancies have been 
extensively studied (Castro Neto et al. 2009, Palacios et al. 2008).
Among other interesting properties, the magnetism 
of  the $\pi$-vacancy model reflects faithfully
the bipartite character of the graphene lattice. For example, 
 the total spin of the system is S=$|N_A-N_B|$, where $N_A$ and $N_B$ are
the number of $\pi$-vacancies in each of the graphene sublattices. 
This behavior can be traced back 
to Lieb's theorem for a half-filled Hubbard model in
a bipartite lattice (Lieb 1989). In the following, we will see that the calculated
data for Co substitutionals and covalently chemisorbed molecules indeed 
follow the predictions of Lieb's theorem. In addition, we analyze in 
detail the spatial decay of the exchange couplings.

\subsection{Magnetic Couplings Between Co Substitutional Impurities in Graphene}
\label{substitutionals-coupling}

\begin{figure}
\includegraphics[width=5.000in]{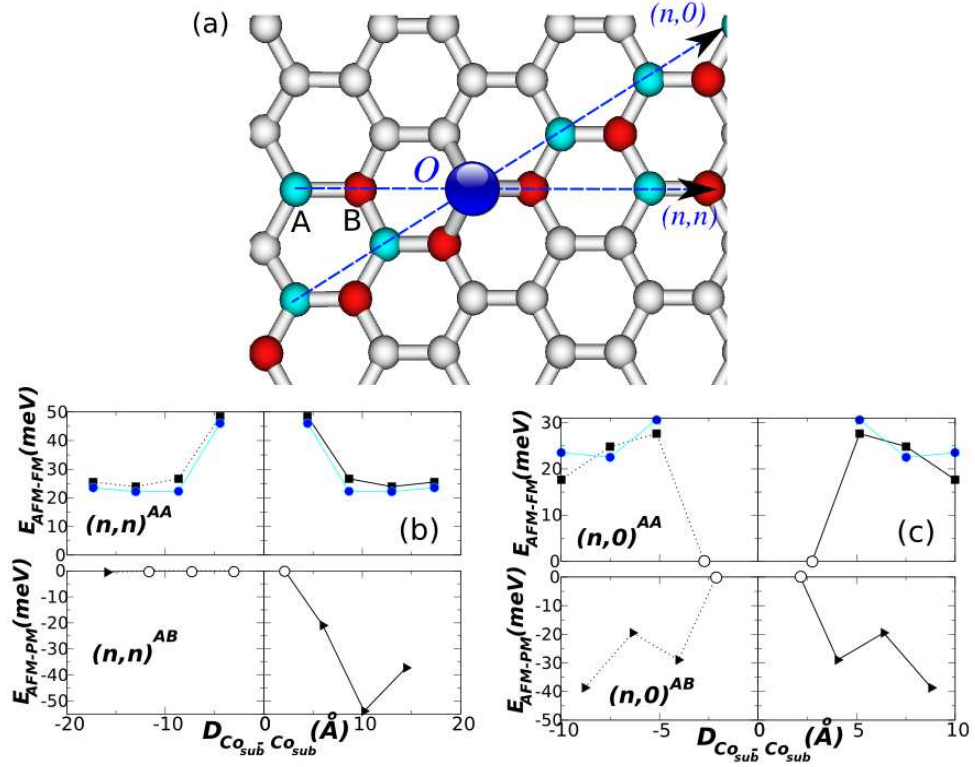}
\caption{(a) Schematic representation  of the geometry
used  to  calculate  the  relative stability  of  
ferromagnetic (FM), antiferromagnetic (AFM)  and spin  compensated  
(PAR) solutions as  a
function of the positions of two Co$_{sub}$ impurities.  Sublattices A
and B  are indicated by  
blue and red circles, respectively.   
One of
the impurities is fixed at a central A-type site, whereas the other is
moved along  the $(n,n)$ and $(n,0)$ directions.
Panels (b) and (c) show the results of the energy differences 
for $(n,n)$ and $(n,0)$ configurations, respectively.   
Solid squares at positive values indicate FM spin alignments, while 
solid triangles at
negative
values
correspond to AFM ones. 
The empty
circles represent spin  compensated solutions and the full circles for AA substitutions
correspond to a fit 
of
 a Heisenberg model (see text for details). 
Adapted from (Santos et al. 2010a).}
\label{fig2-co}
\end{figure}

Here, we consider 
the  magnetic couplings between Co$_{sub}$ defects. 
For this purpose we perform calculations using
a  large  8$\times$8  supercell  with two  Co$_{sub}$  impurities.  We
calculate the energy difference between spin alignments as a function of 
the relative position of  the defects. Figure~\ref{fig2-co}
shows  the  results  along  with  a schematic  representation  of  our
notation. Positive values indicate 
ferromagnetic (FM) spin alignment while negative 
values are 
antiferromagnetic (AFM) ones. Several observations from spin couplings in Fig.~\ref{fig2-co}
can be 
made: ({\it i})  when the impurities  are located in  the same
sublattice (AA  systems) the FM  configuration is more stable  than the 
AFM  one; ({\it  ii}) if  the Co atoms are in opposite sublattices (AB systems) 
it is very difficult to reach  a FM solution,~\footnote{When we could stabilize 
a FM  solution, it lies at higher energy, around 0.2~eV above the PAR one.} 
instead the  system finds either  a spin-compensated  
(PAR) or  an AFM  solution; 
({\it  iii}) at short  distances ($<$  3.0~\AA) the  systems always  converge  to spin
compensated solutions.

\begin{figure}
\includegraphics [width=3.800in]{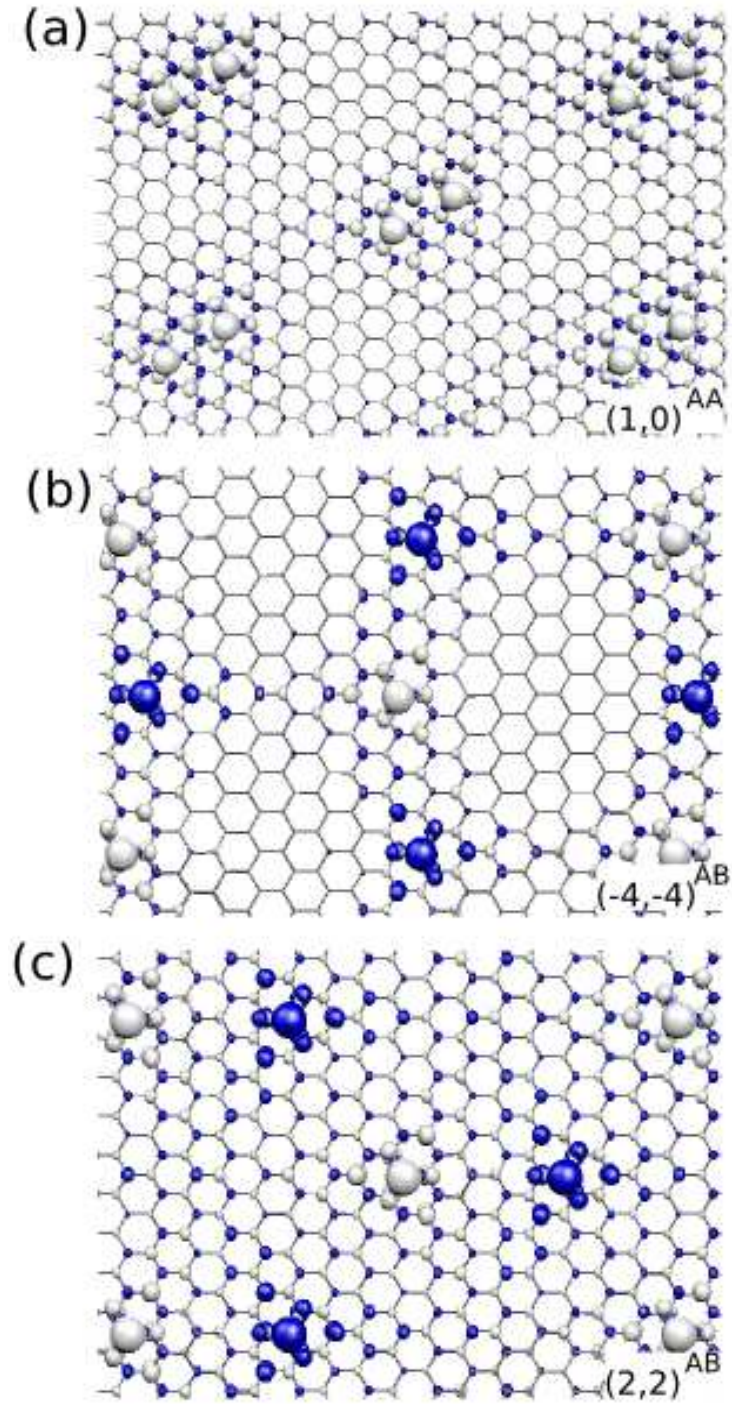}\centering
\caption{(a) Spin  densities for  configurations (a)
$(1,0)^{AA}$, (b) $(-4,-4)^{AB}$  and (c)  $(2,2)^{AB}$(see Fig.~\ref{fig2-co}(a)  for the
nomenclature).  Positive and negative  spin densities are indicated by
light (gray)  and dark (blue) isosurfaces  corresponding to $\pm$0.001
e$^-$/Bohr$^3$, respectively. Adapted from (Santos et al. 2010a).}
\label{fig3-co}
\end{figure}

The FM  cases of Fig.~\ref{fig2-co} always have total spin magnetization about 2.00 $\mu_{B}$. 
The spin population on  every Co  atom remains
almost  constant $\sim$0.50 $\mu_{B}$ and it is  $\sim$0.30 $\mu_{B}$  
on the three C nearest-neighbors.   In other cases the total  spin is zero.
Thus, the  total spin moment  of the system follows  the equation
$S=|N_{sub}^{A}-N_{sub}^{B}|$,  where N$_{sub}^{A(B)}$ is  the number
of Co$_{sub}$ defects in  the A(B) sublattices.  Our total
moment   is   consistent  with   Lieb's   theorem  for   bipartite
lattices~(Lieb 1989).  This   
finding
supports  the analogy,
presented in Section~\ref{Co-analogy}, 
between  the
electronic structure of Co$_{sub}$  defects and single vacancies in a
simplified $\pi$-tight-binding description of graphene.

Some selected  configurations have their 
spin magnetization densities plotted in Fig.~\ref{fig3-co}.
Although  the spin is 
quite 
localized on
the Co  atom and the neighboring  C atoms, 
part of 
the magnetization density is delocalized with alternated signs in  both graphene sublattices.
The triangular  spin patterns reflect the
three-fold symmetry  of the layer with
different  orientations for A and B substitutions. 
This explains the anisotropic AB interaction
along the  $(n,n)$ direction seen  in Fig.~\ref{fig2-co} (b):  the energy
difference   between   AFM   and   PAR   solutions   for   $(n,n)^{AB}$
configurations  strongly  depends  on  the relative  position  of  the
impurities,  showing  such  a directionality.   
Similar  patterns  have already been observed 
experimentally~(Kelly et al. 1998, Mizes et al. 1989, Rutter et al. 2007, Ruffieux et al. 2000)
for point defects in graphene 
using STM  techniques and 
theoretically discussed for
$\pi$-vacancies~(Yazyev 2008, Palacios et al. 2008, Pereira et al. 2008). For
Co$_{sub}$,  similar STM  experiments should  display the topology 
of the spin densities given in Fig.~\ref{fig3-co}.

We  can also  investigate  the  magnetic  interactions  within  the
framework   of   a   classical   Heisenberg  model: 

\begin{equation}
 H=   \sum_{i<j} J_{AA/AB}({\bf r}_{ij}){\bf  S}_{i}{\bf S}_{j} 
\label{heisenberg}
\end{equation}
\\
where ${\bf S}_{i}$ is  the local  moment  for 
a  Co$_{sub}$  impurity at  site $i$.   The
angular dependence of the exchange 
$J ({\bf r}_{ij})$ is
taken from an analytical RKKY coupling as given in~(Saremi 2007). We fit the 
exponent for the distance decay
to our {\it ab  initio} results.  The exchange
interaction  for AA systems  can be  fitted with  a $|r_{ij}|^{-2.43}$
distance dependence  (see the full  circles in Fig.\ref{fig2-co}  (b) and
(c)).  This distance dependence  is in  reasonable agreement  with the
$|r_{ij}|^{-3}$   behavior  obtained   with   analytical  models   for
substitutional defects  and voids~(Saremi 2007, Vozmediano et al. 2005).  
In the case of AB systems a  simple RKKY-like  treatment fails  to 
describe
accurately
 the interactions, at least for the 
short distances
between defects considered in our calculations.

We 
next explain 
how PAR solutions  
appear
in Fig.~\ref{fig2-co}.
The 
PAR solutions 
are stable
because
defect 
states in  neighboring
impurities 
interact strongly
for AB systems. This   
interaction opens 
an appreciable {\it bonding-antibonding} gap in the 
 $p_z$  defect band.~\footnote{For the  AB systems,  we find  bonding-antibonding 
gaps  
in the 
impurity 
bands ranging from 0.3 eV to 0.9~eV for the $(1,1)^{AB}$
and  the $(-1,-1)^{AB}$ configuration, respectively.  These values are similar to 
the $\sim$0.5~eV spin-splitting of the Co$_{sub}$ defect.  
In fact, all those  AB systems with gaps larger  than 0.4~eV converge 
to PAR solutions.}  
For AA systems, however, the bipartite
character  of  the  graphene  lattice makes  the  interaction  between
the
defects  much  smaller.  This  explains why AA configurations show 
a local  spin polarization.  
Even for AA configurations, 
when the impurities are very close, 
non-magnetic solutions are stabilized because 
a larger defect-defect interaction opens a  large gap.
It is interesting to point 
out that similar behaviors have been observed  for vacancies 
described within a $\pi$-tight-binding 
model (see Section 1.6.1)~(Yazyev 2008, Kumazaki et al. 2007, Palacios et al. 2008).

\subsection{Covalent Functionalization Induces Magnetism: Universal Properties }
\label{covalent-spin_formationNT}

\begin{figure}
\includegraphics[width=3.200in]{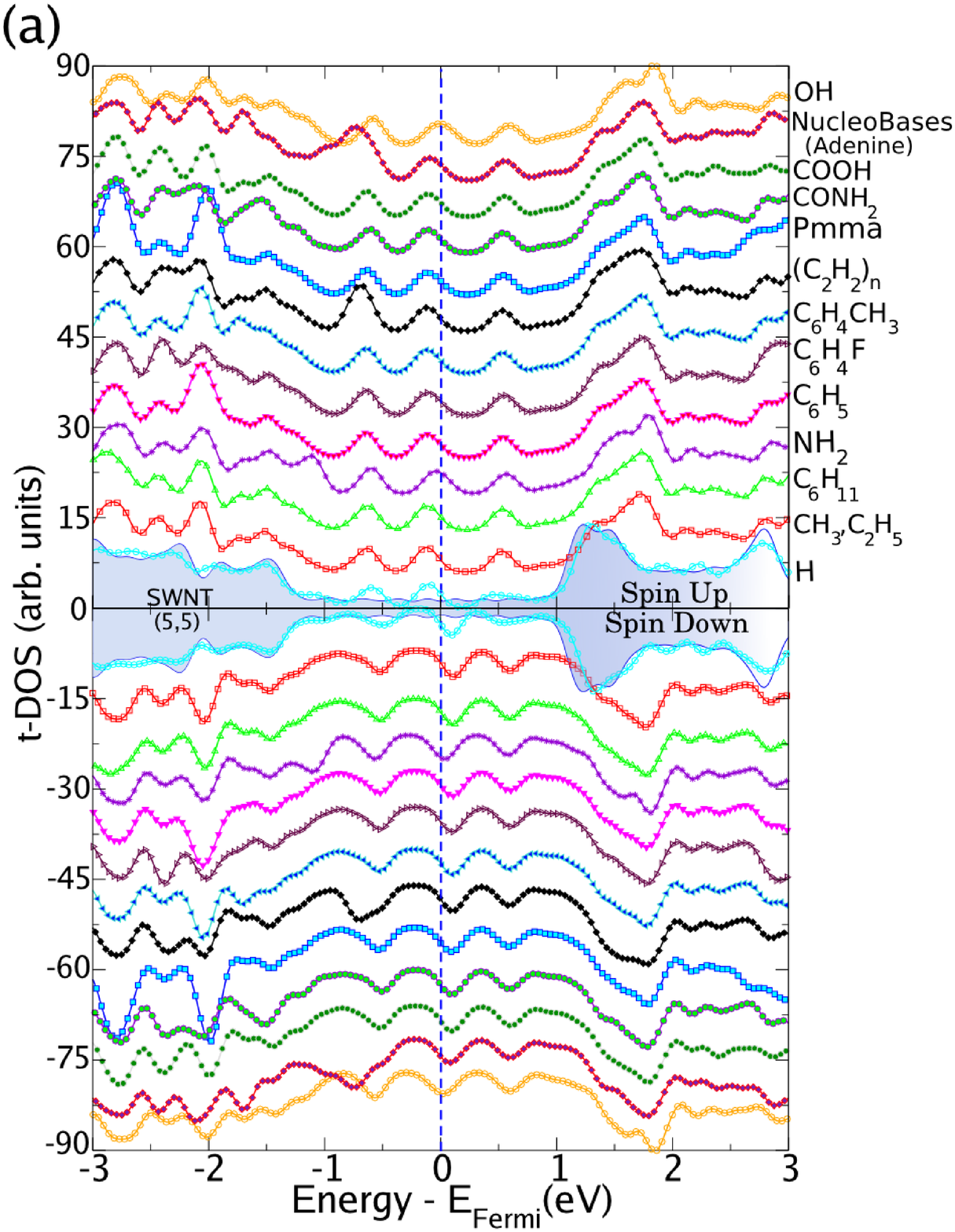} \centering
\includegraphics[width=3.200in]{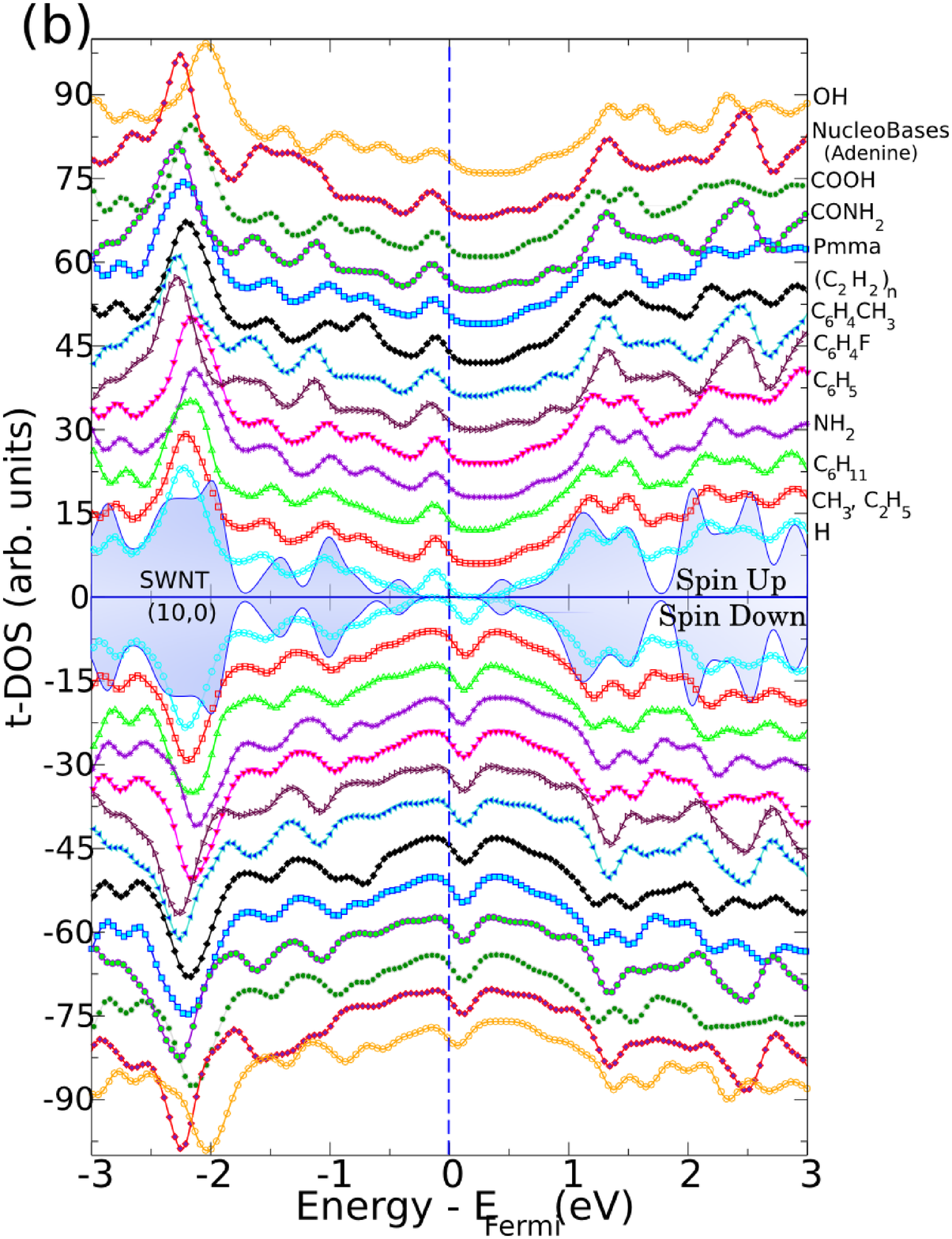}  \centering
\caption{Total spin polarized density of states (t-DOS) for 
(a) (5,5) and (b) (10,0) SWCNTs with a single adsorbate of different types
chemisorbed on top of 
one carbon atom in the supercell. 
Positive and negatives t-DOS correspond to spin up and spin down, respectively.
The t-DOS for pristine (5,5) and (10,0) SWCNTs is also shown for comparison. 
For clarity, the curves in panel (a) and (b) have been shifted and smoothed with a Lorentzian 
broadening of 0.12 eV. Fermi energy is marked by the dashed lines and 
is set to zero in both panels.
}
\label{fig1-adsorbates2}
\end{figure}

In this Section we show that, apart from playing an increasingly important
role 
in  technological applications, 
chemical functionalization can be also used to induce spin moments 
in carbon nanostructures.
 Here, we focus on SWCNT and demonstrate that, 
when a single C-C bond is established 
on
 the carbon surface by 
covalent functionalization, a spin 
moment is induced 
into
the system. 
This moment has 
a universal value of 1.0 $\mu_B$ and is independent 
of 
the particular adsorbate. 
In our recent work (Santos et al. 2011) 
we showed that this effect occurs for a wide class of organic and inorganic molecules of 
different biological and chemical activity (e.g. alkanes, polymers, diazonium 
salts, aryl and alkyl radicals, nucleobases, amido and amino groups, acids).
Furthermore, we have recently found that a similar universal behavior 
is obtained for covalent functionalization of graphene (Santos et al. 2012b).
We have also found that, either for metallic and semiconducting SWCNTs 
or for graphene, 
only when 
neighboring adsorbates are located 
at the same sublattice a spin moment 
is developed.  
For metallic tubes and graphene the local moments align ferromagnetically, 
while for
semiconducting tubes we have almost 
degenerate 
FM and 
AFM spin solutions~(Santos et al. 2011, Santos et al. 2012b).

\begin{figure}
\includegraphics[width=4.8400in]{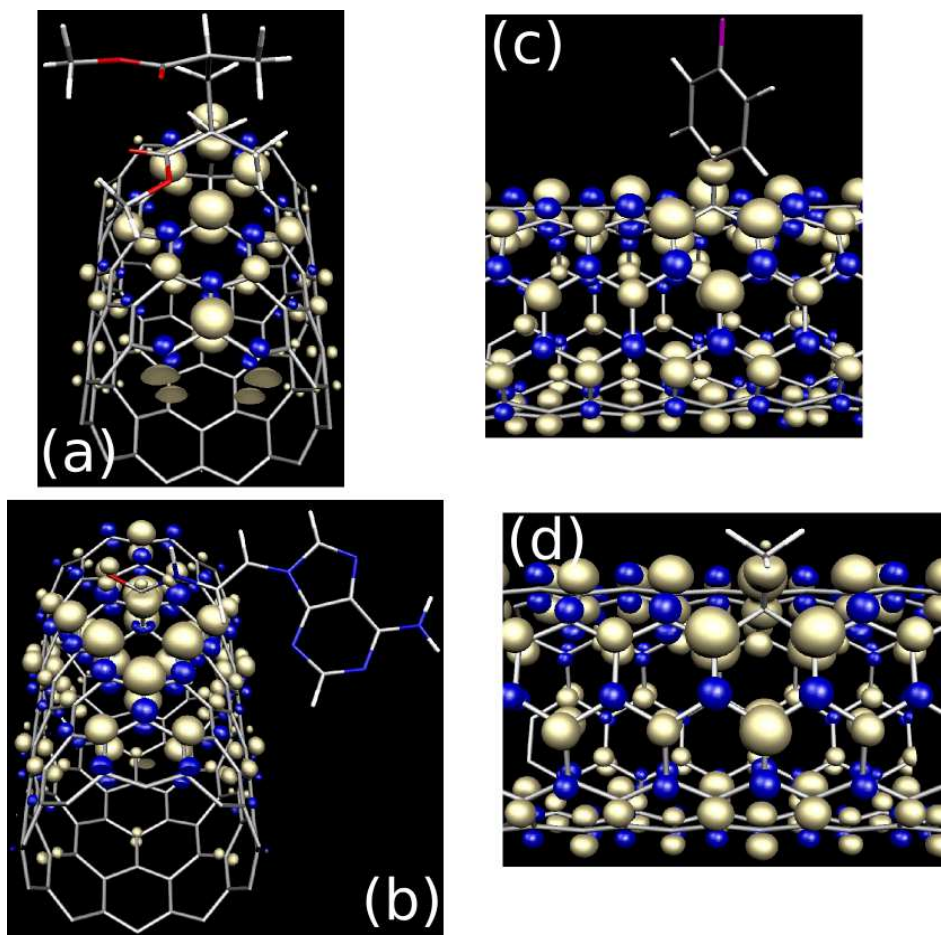}\centering
\caption{Isosurface of the magnetization density induced by some adsorbates at the  
SWCNT surface: (a) Pmma and (b) Adenine 
derivative in a (10,0)
tube; and (c) C$_{6}$H$_{4}$F and (d) CH$_{3}$ in a (5,5)
tube.
Majority and minority spin densities correspond respectively to light and  dark  surfaces, which 
alternate  on the honeycomb lattice with a long decaying order in all cases. 
The cutoff is at 
$\pm$0.013  $e^-/bohr^3$. 
Adapted from (Santos et al. 2011).
}
\label{fig2-tube-adsorbate}
\end{figure}

To understand the origin of the 
spin moment induced when a covalent 
bond is formed 
in
the tube wall, 
we analyze the total spin-polarized density of states (t-DOS) when 
different adsorbates are attached to metallic and semiconducting tubes
(see Figure \ref{fig2-tube-adsorbate} for the structure of some 
of these systems). 
Figure \ref{fig1-adsorbates2} presents results for (a) (5,5) and (b) (10,0) SWCNT's. 
We consider first the well-known case of the adsorption of atomic hydrogen.
In both cases, when a single H atom chemisorbs on top of a C atom
a defect state appears pinned at E$_{F}$ 
with full spin polarization. This state is mainly 
composed of the $p_z$ orbitals 
at 
the nearest C neighbors of the
defect site, with almost no contribution from the adsorbate. 
A detailed Mulliken analysis 
of this 
$p_z$-defect state assigns a contribution 
of the adsorbate of about $\sim$1$\%$.  
Thus, the
adsorbate has a primary role in creating the bond with the nanotube, 
and the associated defect level, but it does not appreciably contribute to the spin moment.
More complex adsorbates, notwithstanding their biological and chemical activity 
(e.g. alkanes, polymers, diazonium salts, aryl and alkyl radicals, nucleobases,
amido and amino groups, acids), show a similar behavior. This is observed 
in the 
t-DOS curves corresponding to other adsorbates
in metallic (5,5) and 
semiconducting (10,0) SWCNTs as shown in Figure \ref{fig1-adsorbates2}(a) and 
\ref{fig1-adsorbates2}(b), respectively. Several common points are 
worth mentioning:
(i) All molecules induce a spin moment of 1.0~$\mu_{B}$ 
localized at the carbon surface;
(ii) The origin of the spin polarization corresponds 
to a $p_{z}$-defect state
with a character and a spatial distribution
similar to those of the state appearing at E$_F$ for
a $\pi$-vacancy defect;
(iii) The t-DOS around E$_{F}$ follows the same pattern in all cases. 
This match demonstrates that the spin moment 
induced by 
covalent functionalization is 
largely
independent of 
the particular type of adsorbate. These results also
demonstrate the complete analogy 
between a single C-H bond 
and 
more complex adsorbates linked
to graphene through a single C-C bond 
(or other weakly-polar covalent 
bonds)
 (Santos et al. 2011, Santos et al. 2012b).
Such similarity is not
obvious and could not be easily anticipated.

Next, we study the spin polarization texture induced by the adsorbates
on the carbon nanotube wall. The analysis of local 
spin moments for all the adsorbates 
assigns general trends to both types, metallic and 
semiconducting, of SWCNTs. The C atoms that participate directly in the 
bond formation, at either the molecule or the surface, show a local spin moment smaller 
than $\sim 0.10 \mu_{B}$. However, the wall carbon atoms contribute with 0.40 $\mu_{B}$ 
in the three first C nearest-neighbors, -0.10 $\mu_{B}$ 
in the next nearest-neighbors, 0.20 $\mu_{B}$ in the third-neighbors.
The adsorbate removes a $p_{z}$ electron from the adsorption site, 
and leaves the $p_z$ states of the 
nearest carbon neighbors uncoordinated. This gives rise to a 
defect state localized in the carbon layer, reminiscent of that 
of a vacancy in a $\pi$-tight-binding model of graphenic nanostructures.
The carbon spins polarize parallel 
(antiparallel) respect to the C atom that binds to the surface 
when sitting in the opposite (same) sublattice.
Figure \ref{fig2-tube-adsorbate} shows the magnetization density
in semiconducting (10,0) and metallic (5,5) SWCNTs for several molecules:
(a) Pmma polymer chain~(Haggenmueller et al. 2000), 
(b) Adenine group nucleobase~(Singh et al. 2009), (c) C$_6$H$_4$F salt~(Bahr et al. 2001) and 
(d) CH$_3$ molecule~(Saini et al. 2002). 
The spin density 
in the metallic (5,5) (Figure \ref{fig2-tube-adsorbate}(c) 
and \ref{fig2-tube-adsorbate}(d)) is more spread over the whole surface 
than in the semiconducting (10,0) (Figure \ref{fig2-tube-adsorbate}(a) 
and \ref{fig2-tube-adsorbate}(b)). 
Thus, the electronic character 
of the nanotube wall plays a role in determining the localization 
of the defects states and,
as will be seen below, in mediating the interaction between adsorbates.

\begin{figure}
\includegraphics[width=5.5200in]{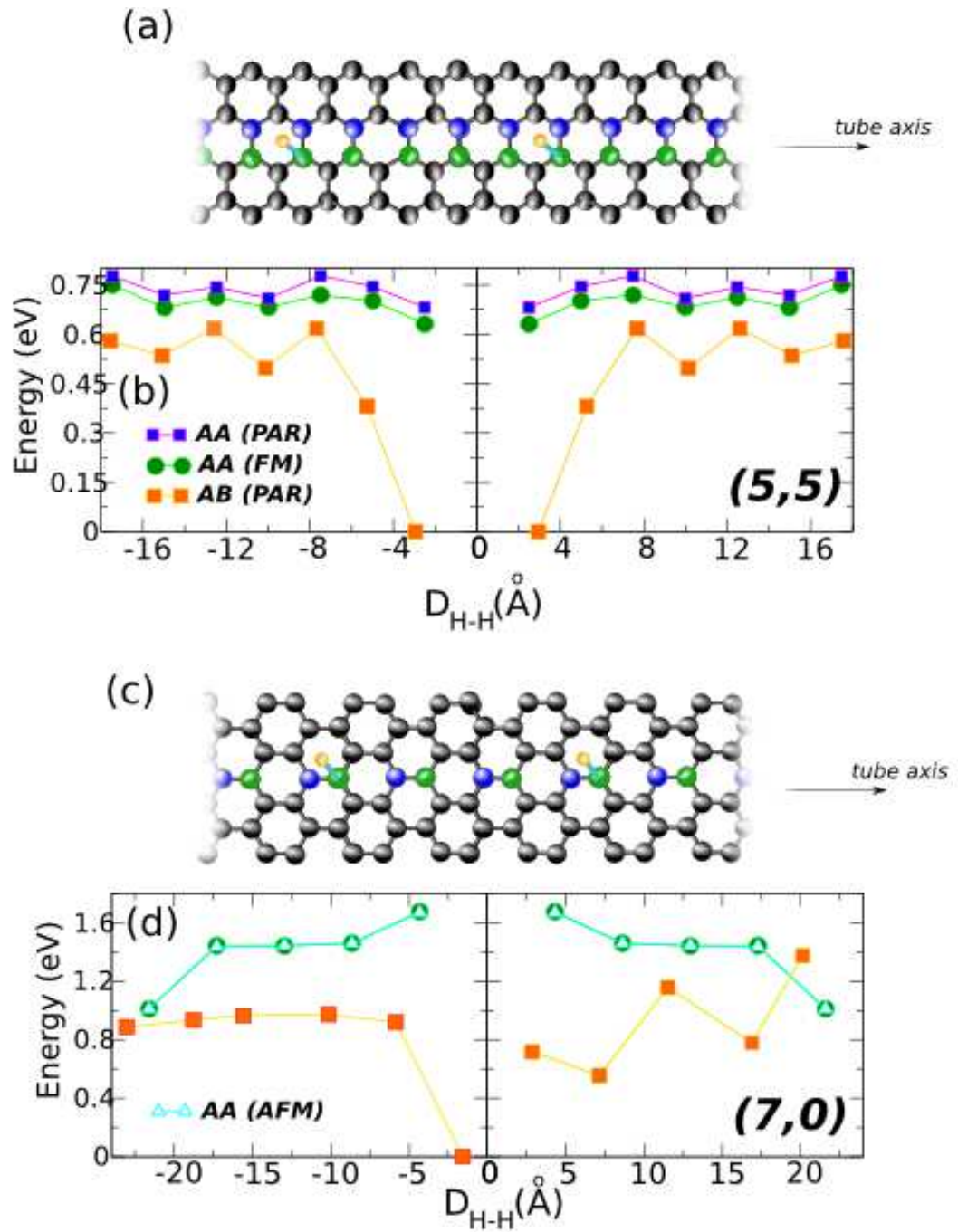}\centering
\caption{Variation of total energy 
as a function of 
the relative adsorption positions of two H  atoms on (a) a (5,5) and (c) a (7,0) SWCNT.
for different magnetic solutions.
One of the 
adsorbates moves 
parallel to the axis of the tube,
while the other 
remains at the origin.
In (b) and (d), the light (yellow) and dark (blue) squares correspond to PAR
spin solutions at AB and AA sublattices, respectively;
the circles and triangles indicate the FM and AFM 
solutions, respectively, at the same sublattice. Adapted from (Santos et al. 2011).
}
\label{fig3-tube-adsorbate}
\end{figure}

Now we deal with the 
relative stability of the different magnetic solutions 
when two molecules are simultaneously adsorbed on the walls of CNTs.
Due to the universal character of the magnetism associated 
with covalent functionalization of SWCNTs, and in order
to alleviate the computational effort, we have considered here hydrogen atoms.
However, we have explicitly checked for some configurations 
for
the case of SWCNTs, and
for flat graphene (Santos et al. 2012b), that identical results
are obtained when using CH$_3$ instead of H. 
For the metallic (5,5) and semiconducting (7,0) single wall CNTs, we calculate 
the variation of the total energy for several spin alignments 
as a function of the distance between the adsorbates. 
The relative 
positions
of the adatoms along the tube are schematically shown in 
Figures \ref{fig3-tube-adsorbate}(a) and \ref{fig3-tube-adsorbate}(c). 
One H  is  sited  at the  origin, and the other sites in  different positions along of the tube axis.
Several observations can be first made on the stability when
two adsorbates are located  at the  same sublattice (AA configurations).
In the metallic (5,5), the FM configuration is most stable than the non-magnetic one (PAR). 
The energy difference between these two spin solutions along the tube axis oscillates 
and no AFM solution could be stabilized at all. In the semiconducting (7,0), the FM and 
AFM solutions are almost  degenerate, 
with a small energy difference (exchange coupling).

If the two molecules are now located at different sublattices 
(AB configurations), we were not able to stabilize any 
magnetic solution for both nanotubes.
Instead the systems 
are
more stable without 
magnetic polarization. 
This behavior for adsorbates at opposite sublattices 
is related to the interaction between the defect levels.  
As already pointed out for Co substitutionals, 
while 
for AA configurations the interaction is negligible, for AB ones 
this interaction opens a bonding-antibonding
gap around $E_{F}$ in the $p_z$ defect band and, thus, contributes to the 
stabilization of PAR solutions. If the gap is larger than the spin splitting of the majority 
and minority spin defect bands the system will be non-magnetic (see Section \ref{substitutionals-coupling}).
In fact, our detailed analysis of the band structure  
fully confirmed 
such
an
explanation. However, it is worth 
noting that AB adsorption seems to be always more stable in 
our calculations. This indicates that if the adsorption takes place at 
random sites, the magnetic solutions will only be stable for low density functionalization.

\section{Conclusions}

In this Chapter we have reviewed the structural, electronic 
and magnetic properties of 
two types of defects, substitutional metal impurities and 
$sp^3$-type covalent
functionalization,  
in carbon nanostructures. We
have focused on their role 
to induce and control magnetism in graphene and carbon nanotubes. 
Density functional theory was the main tool used to compute 
the properties of the studied systems. We also
developed simple models to understand the observed trends. 
For instance, substitutional dopants in graphene were understood in terms of 
the hybridization of the $d$ states of the metal atoms with 
those of an unreconstructed carbon vacancy. 
The main ingredients of the model are the assumption of a three-fold symmetric 
bonding configuration and the approximate knowledge of the relative energy 
positions of the levels of the carbon monovacancy and the $d$ shell of the metal
impurity as we move along the transition series. With this model, we understood 
the variations of the electronic structure, 
the size and localization of the spin moment, and the binding energy 
of transition, noble metals and Zn substitutional impurities in 
graphene (Santos et al 2010b). 
Our model also allowed us to draw an analogy between 
substitutionals of the late transition metals
and the symmetric D$_{3h}$ carbon vacancy. 

As a result of our analysis, a particularly powerful analogy was established between 
the substitutional Co impurity and the fictitious $\pi$-vacancy in graphene (Santos et al 2010a).
The $\pi$-vacancy corresponds to a missing $p_z$ orbital in a simple description of graphene
using a $\pi$-tight-binding model. The magnetic properties of the $\pi$-vacancies 
have been 
extensively studied. 
This analogy 
brings our results for the magnetism of Co$_{sub}$ defects into contact with
the predictions of Lieb's theorem for a half-filled Hubbard model in
a bipartite lattice. We found that, according to
Lieb's theorem, the total spin of the system is S=$|N_A-N_B|$, where $N_A$ and $N_B$ are
the number of substitutions performed in each of the graphene sublattices. 
Thus,
the couplings between local moments for 
AA substitutions are 
ferromagnetic and
predominantly antiferromagnetic for AB substitutions. We have also used a simple RKKY-model
to extract the distance decay of the couplings. 

Adsorbates attached to graphene or SWCNTs through covalent bonds, particularly
if the bonds are weakly polar, constitute another example of defects
whose magnetism is analogous to that of the $\pi$-vacancy. 
We have analyzed the magnetic properties 
induced by such 
a
covalent functionalization using many types of adsorbates: 
polymers, diazonium salts, aryl and alkyl radicals, nucleobases, amide and amine 
groups, sugar, organic acids, for SWCNTs (Santos et al. 2011) and graphene (Santos et al. 2012b). 
A universal spin moment of 1.00 $\mu_{B}$ is induced 
on
the 
carbon surface 
when
a single C-C bond is formed between an adsorbate and the graphenic layer.
In metallic carbon nanotubes and graphene, 
molecules 
chemisorbed at the same sublattice (AA adsorption) 
have their
local moments 
aligned ferromagnetically. 
In semiconducting nanotubes, FM and AFM solutions are almost degenerate even for AA adsorption.
For two molecules in different sublattices (AB adsorption), 
we could not stabilize any magnetic solution, and the system is more 
stable without a local spin moment.

We have also explored the possibility to control the magnetism induced by
substitutional impurities using 
mechanical deformations. 
We have found that 
the spin moment of substitutionally Ni-doped graphene 
can be controlled by applying mechanical deformations that break the hexagonal 
symmetry of the layer, like curvature or uniaxial strain. Although
Ni$_{sub}$ impurities are non-magnetic in flat graphene,
we have observed that stretching the layer by a few percents along different 
crystalline directions is enough to turn the non-magnetic ground state of 
Ni atoms embedded in graphene to a magnetic state (Santos et al. 2012a). 
The spin moment slowly increases as a function of the applied strain. 
However, at a critical strain value of 6.8\%, 
a sharp transition to high spin ($\sim$2~$\mu_B$) state is observed.
This transition is   
independent 
of
the orientation 
of 
the applied strain. A detailed 
analysis indicates that this strain-tunable spin moment is the result of 
changes 
of 
the positions of three defect levels around Fermi energy which are antibonding 
combinations of the Ni $3d$ states and the $2p_{z}$ and $2sp^2$ orbitals of the neighboring C atoms. 
This tunable magnetism observed in Ni$_{sub}$ defects via 
strain control may play an interesting role in flexible spintronics devices.

Our calculations show that Ni$_{sub}$ magnetism can also be switched on 
by applying curvature (Santos et al. 2008).  
For metallic carbon 
nanotubes the curvature of the carbon layer around the defect 
can drive the transition of the Ni$_{sub}$ impurities to a magnetic state.
For semiconducting tubes, the Ni$_{sub}$ impurities remain non-magnetic irrespective of the tube
diameter. 
We have analyzed in detail the origin and distribution of the magnetic moment. 
We found that 
the spin moment associated 
with Ni$_{sub}$ impurities 
forms
accompanied by the simultaneous polarization of delocalized
electronic states in the carbon layer. Furthermore, the spin 
moment of Ni$_{sub}$ is a signature of the metallicity of 
the structure: only metallic tubes develop a moment that depends on the tube diameter and Ni concentration.

Our work predicts a complex magnetic behavior
for transition metals impurities in carbon nanotubes and graphene. 
This investigation is highly relevant 
in
the interpretation of experimental
results since 
it has been proposed that appreciable amounts
of metal atoms could be incorporated into the carbon network,
forming this type of substitutional defects in the course of 
synthesis, and are very difficult to eliminate afterwards.
Our results also indicate that
covalent functionalization provides 
a powerful route to tune the magnetism of graphene and carbon nanostructures. 
This is particularly attractive due to the recent 
successful synthesis
of different graphene derivatives using surface chemical 
routes ( Cai et al. 2010, Treier et al. 2010). 
Thus, the synthesis of carbon nanostructures with functional groups
at predefined positions, for example starting from previously functionalized
monomers, seems plausible nowadays. According to our results, this could   
be applied to synthesize magnetic derivatives of
graphene that behave according 
to well studied theoretical models 
like the $\pi$-vacancy.

\end{document}